\documentclass{emulateapj}

\newcommand{\lya}{Ly$\alpha$} 

\begin{document}
\title{Spectroscopy of \textit{z} $\sim$ 6 $i$-dropout Galaxies:
Frequency of $Ly\alpha$ Emission and the Sizes of $Ly\alpha$ Emitting
Galaxies\altaffilmark{1,}\altaffilmark{2,}\altaffilmark{3}}
\altaffiltext{1}{Based on observations taken with the W.M. Keck Observatory, which is operated as a scientific partnership among the California Institute of Technology, the University of California, and NASA.}
\altaffiltext{2}{Based on observations made with the NASA/ESA \textit{Hubble Space Telescope}, which is operated by the Association of Universities for Research in Astronomy, Inc., under NASA contract NAS 5-26555.  These observations are associated with programs 7817, 9270, 9301, 9583, and 9803.}
\altaffiltext{3}{Based on observations collected at the European Southern Observatory, Paranal, Chile (LP166.A-0701 and 169.A-045).}

\author{C. C. Dow-Hygelund\altaffilmark{4}}

\author{B.P. Holden\altaffilmark{5}}

\author{R.J. Bouwens\altaffilmark{5}}

\author{G.D. Illingworth\altaffilmark{5}}

\author{A. van der Wel\altaffilmark{6}}

\author{M. Franx\altaffilmark{7}}

\author{P.G. van Dokkum\altaffilmark{8}}

\author{H. Ford\altaffilmark{6}}

\author{P. Rosati\altaffilmark{9}}

\author{D. Magee\altaffilmark{5}}

\author{A. Zirm\altaffilmark{6}}

\altaffiltext{4}{Physics Department, University of California,
    Santa Cruz, CA 95064; cdow@scipp.ucsc.edu}
\altaffiltext{5}{Department of Astronomy and Astrophysics, University of California,
    Santa Cruz, CA 95064; holden@ucolick.org, bouwens@ucolick.org, gdi@ucolick.org, magee@ucolick.org}
\altaffiltext{6}{Department of Physics and Astronomy, Johns Hopkins University,  3400 North Charles Street
Baltimore, MD 21218-2686; wel@pha.jhu.edu, ford@pha.jhu.edu, azirm@pha.jhu.edu}
\altaffiltext{7}{Leiden Observatory, P.O. Box 9513, J. H. Oort Building, Niels Bohrweg 2, NL-2300 RA, Leiden, Netherlands; franx@strw.leidenuniv.nl}
\altaffiltext{8}{Department of Astronomy, Yale University, P.O. Box 208101, New Haven, CT 06520-8101; dokkum@astro.yale.edu}
\altaffiltext{9}{European Southern Observatory, Karl-Schwarzchild-Strasse 2, D-85748 Garching, Germany; prosati@eso.org}

\shorttitle{Spectroscopic Confirmation of \textit{z} $\sim$ 6 LBGs}

\begin{abstract}

We report on deep spectroscopy using LRIS on Keck I and FORS2 on the
VLT of a sample of 22 candidate \textit{z} $\sim$ 6 Lyman Break
galaxies (LBGs) selected by the $i_{775}-z_{850}>$ 1.3 dropout
criterion.  Redshifts could be measured for eight objects.  These
redshifts are all in the range \textit{z} = 5.5 - 6.1, confirming the
efficiency of the $i_{775}-z_{850}$ color selection technique.  Six of
the confirmed galaxies show \lya\ emission.  Assuming that the 14
objects without redshifts are \textit{z} $\sim$ 6 LBGs, but lack
detectable \lya\ emission lines, we infer that the fraction of \lya\
emitting LBGs with \lya\ equivalent widths greater than 20 \AA\ among
\textit{z} $\sim$ 6 LBGs is $\approx$30$\%$, similar to that found at
\textit{z} $\sim$ 3.  Every \lya\ emitting object in our sample is
compact with r$_{hl}$ $\leq$ 0\farcs14.  Furthermore, all the \lya\
emitting objects in our sample are more compact than average relative
to the observed size-magnitude relation of a large \textit{i}-dropout
sample (332 candidate \textit{z} $\sim$ 6 objects).  We can reject the
hypothesis that the \lya\ emitting population is a subset of the rest
of the \textit{z} $\sim$ 6 LBG population at $>$97$\%$ confidence.  We
speculate the small sizes of \lya\ emitting LBGs are due to these
objects being less massive than other LBGs at \textit{z} $\sim$ 6.

\end{abstract}


\section{Introduction}

Distinguishing high redshift galaxies from interlopers at lower
redshifts can be a challenging process \citep{Stern1999}.  One
particularly unique trait of high-redshift galaxies is the continuum
break they possess as a result of absorption by neutral hydrogen along
the line-of-sight \citep{Madau1995,Dickinson1999}.  Putting together
this feature with the particularly blue colors of star-forming
galaxies, we have a simple but efficient technique for selecting
high-redshift star-forming galaxies \citep{Steidel1995,Dickinson1999}.
This technique has been extensively tested through spectroscopy on
numerous Lyman break galaxies (LBGs) at \textit{z} $\sim2.5-4.5$
\citep{Steidel1999} and an increasing number of galaxies and quasars
at \textit{z} $\gtrsim$ 4
\citep{Weymann1998,Spinrad1998,Dey1998,Lehnert2003,Nagao2004,Fan2001,Bunker2003,Dickinson2004,Vanzella2006}.
In addition, there are now surveys that exploit narrowband excess from
\lya\ emission \citep[see][]{Taniguchi2003b} to compile large ($>20$)
samples of \lya\ emitters (LAEs) at \textit{z} $\approx$ 5.7
\citep{Rhoads2003,Hu2004,Shimasaku2005} and at \textit{z} $\approx$
6.6 \citep{Kashikawa,Stern2005}.

Several groups, utilizing these techniques, have obtained
statistically significant samples of galaxies at \textit{z} $\sim$ 6
\citep{Rhoads2001,Ajiki2003,Bouwens2003,Bouwens2004a,Bouwens2005,Stanway2003b,Dickinson2004,Ajiki2005},
as well as identifying candidate galaxies out to \textit{z} $\sim$ 7-8
\citep{Hu2002,Kodaira2003,Bouwens2004b,Taniguchi2004,Stern2005,Rhoads2004a,Kneib2004,Bouwens2006,Iye}.
This has led to great progress in our understanding of the early
universe, providing us with constraints on both the epoch of
reionization \citep[see][]{Loeb2001,Fan2003,Malhotra&Rhoads2004} and
the evolution of the global star formation rate (SFR) density
\citep{Bouwens2003,Bouwens2004a,Bouwens2005,Hopkins2004,Giavalisco2004}.
However, to ensure that these results are accurate, it is essential
that we understand nature of the galaxy populations being selected by
these techniques and can quantify their relation to the underlying
population of star-forming galaxies at $z\gtrsim6$.  This is paramount
if we are to construct a self-consistent picture of the galaxy
population from the different selection techniques.  Issues that need
to be addressed include (1) quantifying the distribution of \lya\
equivalent widths for star-forming galaxies at $z\sim6$, (2)
determining the impact of these results on narrowband and broadband
selections, and (3) using search results to construct a
self-consistent picture of the galaxy population at $z\gtrsim6$.

In this study, we utilize spectroscopy to study the nature of $z\sim6$
$i_{775}$-dropouts identified by \citet{Bouwens2003,Bouwens2004a} in
deep ACS imaging data.  By contrasting the properties of our
spectroscopic sample with similar selections at $z\sim3$, we examine
the evolution of the spectral properties of these sources with
redshift.  In addition, we pay particular attention to the
implications of these results for future studies of high redshift
galaxies using either narrowband or broadband selection techniques --
noting their respective strengths and complementarity.  A brief
outline of this paper follows.  A summary of the \textit{i}-dropout
sample and spectroscopic observations is presented in \S2.  In \S3 we
present the spectra and their interpretations.  In \S4, we compare our
results with those obtained with narrowband selections at \textit{z}
$\sim$ 6.  In \S5, we discuss the fraction of LAEs in the
\textit{i}-dropout sample, as compared with lower redshift samples.
We also investigate a possible link between the morphologies of the
\textit{i}-dropouts and their \lya\ emission.  Finally, in \S6 we
summarize our findings, and discuss their implications for future
high-redshift surveys.

% fix sections stuff!!!!!

In this paper we adopt the 'concordance' cosmology: a
$\Lambda$-dominated, flat universe with $\Omega_{M}=0.3$,
$\Omega_{\Lambda}=0.7$, and $H_{o}$=71 km s$^{-1}$ Mpc$^{-1}$
\citep{Bennett2003}.  All magnitudes are given in the AB system
\citep{Oke&Gunn1983}.  We denote the \textit{Hubble Space Telescope}
Advanced Camera for Surveys (ACS) F435W, F606W, F775W, and F850LP
passbands as $B_{435}$, $V_{606}$, $i_{775}$ and $z_{850}$
respectively.

\section{Sample + Observations}

\subsection{Optical Broadband Dropout-Selected Survey} 

Our spectroscopic sample was derived from \textit{HST} ACS
observations over two separate fields: RDCS 1252.9-2927 (CL1252), a
cluster at \textit{z} = 1.237 selected from the ROSAT Deep Cluster
Survey \citep{Rosati1998,Rosati2004}, and the Ultra Deep Field
parallel fields \citep[UDF PFs;][] {Bouwens2004a}. A 6\arcmin\ by
6\arcmin\ mosaic of ACS Wide Field Camera $i_{775}$ and $z_{850}$ band
images was acquired around CL1252, with 4\arcmin\ by 4\arcmin\ VLT
ISAAC IR imaging of the mosaic center \citep{Lidman2004}.  The UDF PFs
were imaged with a 4\farcm5 by 4\farcm5 ACS mosaic in the $B_{435}$,
$V_{606}$, $i_{775}$ and $z_{850}$ band filters.  For further details
concerning the imaging data and our photometry over these fields, we
refer the reader to \citet{Bouwens2003,Bouwens2004a}.

The $i_{775}$ and $z_{850}$ band imaging data over these fields allow
us to readily select LBGs in the range 5.5 $<$ \textit{z} $<$ 6.2
\citep{Bouwens2003} using an $i$-dropout criteria.  In the case of
CL1252, the selection was based upon a $i_{775}-z_{850}>$ 1.3 color
cut and for the UDF PFs \citep{Bouwens2004a}, a $i_{775}-z_{850}>$ 1.4
cut plus a null detection in $V_{606}$ (2 $\sigma$).  Note that the
colour selection we use for following up sources in the CL1252 field
is more inclusive than the $i_{775}-z_{850}>$ 1.5 criterion used in
\citet{Bouwens2003}.  The $i_{775}-z_{850}>$ 1.3 criterion used for
selection in the CL1252 field resulted from a small systematic error
that was present in the early $z_{850}$ ACS zeropoint.  The typical
error for the $i_{775}-z_{850}$ colors is 0.4 mag.  In the case of a
non-detection, the $i_{775}$ flux was set to the 2$\sigma$ upper
limit.  The error and upper limits are incorporated into all
simulations that determine selection volume for the different samples.
A total of 25 \textit{i}-drops were identified in the CL1252 field,
and 40 over the UDF PFs, with $10\sigma$ limiting magnitudes of
$z_{850}$ = 27.3 and 27.8 respectively.  This corresponds to surface
densities of 0.5 $\pm$ 0.1 objects arcmin$^{-2}$ for CL1252, and 1.4
$\pm$ 0.2 objects arcmin$^{-2}$ for the UDF PFs.

\begin{deluxetable*}{cccccccccc}
\tabletypesize{\scriptsize}
\tablecaption{Properties of our RDCS 1252-2927 Spectroscopic Sample \label{CL1252}}
\tablewidth{0pt}
\tablehead{
\colhead{Object ID} & \colhead{Previous ID\tablenotemark{a}}
&\colhead{RA} & \colhead{Dec} & \colhead{$z_{850}$}  &
\colhead{$J$} & \colhead{$Ks$} &  \colhead{$i_{775}-z_{850}$} &
\colhead{r$_{hl}$(\arcsec)} & \textit{z}}
\startdata
BD38\tablenotemark{b,c} & 1252-5224-4599 &  12 52 56.888 & -29 25 55.50
&  24.3$\pm$0.1 & 24.1$\pm$0.1 & 23.8$\pm$0.1 & 1.5 & 0.29 &5.515\\
BD03\tablenotemark{b,c} & 1252-2134-1498 &  12 52 45.382 & -29 28 27.11
&  25.6$\pm$0.1 & 25.2$\pm$0.4 & $>$25.6 & $>$2.1 & 0.18 &5.980\\
BD27 &    \ldots       &  12 52 51.902 & -29 26 28.60 &  25.6$\pm$0.1 &
26.1$\pm$0.4 & 25.8$\pm$0.4 & 1.5 & 0.11 &\ldots \\
BD58 & 1252-2585-3351 &  12 52 52.283 & -29 28 04.74 &  25.7$\pm$0.1 &
25.6$\pm$0.3 & 25.4$\pm$0.3 & 2.0 & 0.20 & \ldots \\
BD44 & 1252-5058-5920 &  12 53 01.745 & -29 26 03.87 &  25.9$\pm$0.2 &
26.3$\pm$0.4 & 25.5$\pm$0.4  & $>$1.8 & 0.19 & \ldots \\
BD46\tablenotemark{b,d} &    \ldots       &  12 53 05.424 & -29 24 26.22
&  26.1$\pm$0.1  & NA & NA & 1.6 & 0.14 & 5.914\\
BD57 &    \ldots       &  12 52 49.441 & -29 27 53.24 &  26.1$\pm$0.2 &
25.7$\pm$0.3 & 24.9$\pm$0.2 & 1.5 & 0.19 & \ldots\\
BD00\tablenotemark{d} &    \ldots       &  12 52 42.927 & -29 29 20.05 &  26.1$\pm$0.1
& NA    & NA & 1.3 & 0.11 & 5.942\\
BD66 &    \ldots       &  12 52 54.624 & -29 24 56.18 &  26.1$\pm$0.1
& NA    & NA & 1.4 & 0.13  & \ldots\\
BD40 &    \ldots       &  12 52 59.511 & -29 26 58.43 &  26.3$\pm$0.1 &
26.2$\pm$0.3 & 25.9$\pm$0.4 & 1.3 & 0.15 & \ldots\\
BD22 &    \ldots       &  12 52 53.346 & -29 27 10.58 &  26.5$\pm$0.2
& NA    & NA & 1.4 & 0.13 & \ldots\\
BD62 & 1252-3729-4565 &  12 52 56.746 & -29 27 07.63 &  26.7$\pm$0.2 &
26.3$\pm$0.3 & $>$26.2 & $>$1.8 & 0.12 & \ldots\\
BD48 & 1252-3497-809  &  12 52 42.754 & -29 27 18.89 &  27.0$\pm$0.2
& NA    & NA & $>$1.6 & 0.11 & \ldots\\
BD36 &    \ldots       &  12 52 54.944 & -29 25 57.52 &  27.1$\pm$0.1 &
26.3$\pm$0.4 & 25.2$\pm$0.4 & 1.5 & 0.11 & \ldots
\enddata
\tablecomments{Right ascension (hours, minutes, seconds) and
declination (degrees, arcminutes, arcseconds) use the J2000 equinox.
All magnitudes given are AB.  An entry of NA indicates that the object
was outside the ISAAC coverage we had on CL1252 \citep{Lidman2004}.  Each
of these objects were observed for four hours with LRIS.}
\tablenotetext{a}{\citet{Bouwens2003}.}
\tablenotetext{b}{FORS2 22.3 hr integrated spectra also obtained.}
\tablenotetext{c}{\rm{Lacking detectable \lya\ emission; absorption line redshift.}}
\tablenotetext{d}{\rm{\lya\ emission detected; emission line redshift.}}
\end{deluxetable*}

In Table \ref{CL1252} and Table \ref{UDF PFs}, the objects satisfying
the \textit{i}-dropout criteria for spectroscopic follow-up are
listed, along with their $z_{850}$-band, $J$-band and $Ks$-band
magnitudes, and $i_{775}-z_{850}$ colors.  A total of 22 objects were
observed spectroscopically (14 in CL1252 and eight in the UDF PFs).
These objects were randomly selected from the entire
\textit{i}-dropout sample for spectroscopic follow-up, except for the
most luminous dropout in the CL1252 field (BD38) which was explicitly
targetted for follow-up.  In addition, \textit{i}-dropout galaxies
were the primary targets for both the CL1252 and UDF PFs spectroscopic
observations.  The faintest galaxy on our mask had a $z_{850}$-band
magnitude of 27.5.

\begin{deluxetable*}{cccccccc}
\tabletypesize{\scriptsize}
\tablecaption{Properties of our UDF Parallel Spectroscopic Sample \label{UDF PFs}}
\tablewidth{0pt}
\tablehead{
\colhead{Object ID} & \colhead{Previous ID\tablenotemark{a}} &
\colhead{RA} & \colhead{Dec} & \colhead{$z_{850}$} &
\colhead{$i_{775}-z_{850}$} & \colhead{r$_{hl}$(\arcsec)} & \colhead{\textit{z}}}
\startdata
UDF PFs1 i0 &      \ldots        &    03 32 35.604 & -27 57 37.51 &
25.5$\pm$0.1 & 1.9 & 0.25 &\ldots\\
UDF PFs1 i1 &      \ldots        &    03 32 34.892 & -27 57 10.74 &
25.7$\pm$0.1 & 1.5 & 0.11  &\ldots \\
GOODS i6 0\tablenotemark{b}   &      \ldots        &    03 32 39.803 & -27 52 57.88 &
26.1$\pm$0.1 & 1.4 & 0.10  & 5.540\\
UDF PFs i4\tablenotemark{b}       & UDFP1-3851-2438   &    03 32 43.959 & -27 56 43.87 &
26.3$\pm$0.1 & 1.6 & 0.11  &5.857\\
UDF PFs i9       & UDFP1-4650-3354   &    03 32 40.707 & -27 57 24.28 & 26.4$\pm$0.1 & 1.7  & 0.16 & \ldots \\
UDF PFs i1\tablenotemark{b}       & UDFP1-2954-1152   &    03 32 48.368 & -27 55 54.82 & 26.9$\pm$0.2 & 1.5 & 0.09 & 6.005 \\
UDF PFs i2\tablenotemark{b}      & UDFP1-2309-1628   &    03 32 46.410 & -27 55 24.32 & 27.1$\pm$0.1 & $>$2.5 & 0.10  & 6.083\\
UDF PFs i0\tablenotemark{c}    & UDFP1-3407-1028   &    03 32 48.952 & -27 56 16.97 & 27.5$\pm$0.2 & 1.7 & 0.10 & \ldots 
\enddata
\tablenotetext{a}{\citet{Bouwens2004a}.}
\tablenotetext{b}{\rm{\lya\ emission detected.}}
\tablenotetext{c}{This object's LRIS spectrum was contaminated by bleeding from an alignment star spectrum.  Therefore it is excluded from the sample.}
\end{deluxetable*}

\subsection{Keck I LRIS}

We used the Low Resolution Imaging Spectrograph (LRIS) \citep{Oke1995}
on the Keck I 10m telescope for spectroscopic follow-up of all 22
objects in our spectroscopic sample.  A total integration time of
16200 s for CL1252 objects and 7200 s for UDF PFs objects was obtained
using the 600 line mm$^{-1}$ grating blazed at 8500 \AA\ with a
slitlet width of 1\arcsec\ and a minimum slitlet length of 10\arcsec.
The scale ($\Delta\lambda_{res})$ of this configuration is 1.28 \AA\
per pixel.  A series of eight dithered exposures was taken to aid in
cosmic ray removal, and to enhance removal of the fringing in the
near-IR region of the spectra.  The telescope was offset
1\arcsec-3\arcsec along the slits between exposures, which ranged from
1200 to 2250 s in duration.  The observations were taken under
photometric conditions on the nights of UT 2004 February 13-14.

\begin{figure}
\begin{center}
\includegraphics[width=3in, angle=0, scale=1.0]{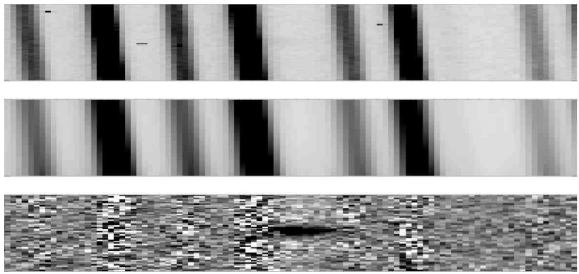}
\caption{\em An illustration of the reduction process for our two
dimensional spectral data.  The upper panel shows the unreduced
spectrum, the middle panel shows a two dimensional fit to the
background, and the lower panel is the reduced spectrum.
\label{bd00_2d}}

\end{center}
\end{figure}

All data reductions were conducted using a slit mask reduction task
developed by Daniel Kelson.  This task yields cleaner background
subtraction and cosmic ray removal than standard slit spectroscopy
procedures \citep{Kelson2003}.  By performing the sky subtraction
before the data have been rectified, this methodology minimizes
reduction errors.  One example of this reduction procedure is
presented in Figure \ref{bd00_2d}.  Wavelength calibration was
performed using the night skylines.  The spectra were flux calibrated
using a sensitivity function derived from observations taken the same
night of the spectrophotometric standard HZ44
\citep{Massey1988,Massey1990}.

\subsection{VLT FORS2}

Three CL1252 \textit{i}-drop objects (BD38, BD03, and BD46) were also
observed with the Focal Reducer/low dispersion Spectrograph 2 (FORS2)
on the 8.2-m VLT YEPUN Unit Telescope in Mask Exchange Unit mode.
These objects were observed as part of a large observing program aimed
at acquiring very deep spectra of both field and cluster elliptical
galaxies in the redshift range 0.6-1.3.  We used the 600z grism with
the OG590 order separation filter, which yielded a scale element of
1.64 \AA\ per pixel. The data presented here were taken from UT 2003
February 28 through March 2 and April 27 through May 24 with a median
seeing of 0\farcs65 and clear conditions. The observations were
carried out in a series of four dithered exposures with equal exposure
times ranging from 14 to 30 minutes each, yielding a total integration
time of 22.3 hours for the sources.  All exposures are added and
weighted such that optimal signal-to-noise ratios were
obtained. Details concerning the data reduction are provided in
\citet{Arjen2005}.

\section{Results}

\subsection{Keck Data}

Six emission line objects were detected out of the observed 22 object
sample.  No other spectral features were found (including continuum)
in the remaining spectra.  Figures \ref{LAE}a - f show our spectra of
these objects, with lines evident at 8440 \AA, 8404 \AA, 8514 \AA,
8609 \AA, 8335 \AA, and 7950 \AA.  Since O {\scriptsize II}
$\lambda$3727 would be resolved into a doublet structure with our
spectral resolution, this interpretation for these lines could clearly
be rejected.  Moreover, line interpretations as H$\beta$
$\lambda$4861.3, H$\alpha$ $\lambda$6562.8, and O {\scriptsize III}
$\lambda$5006.8, 4958.9 could also be discarded due to the lack of
nearby lines and the $i_{775}-z_{850}>$1.3 colors (strong continuum
breaks) of our objects.  For objects BD00, BD46, UDF PFs i1, and UDF
PFs i2, the asymmetry of the line profiles, with absorption on the
blue side, is consistent with absorption by a thick \lya\ forest.
Therefore, it seems quite clear that these emission lines are \lya\
$\lambda$1215.67, placing these objects at \textit{z} = 5.942, 5.9214,
6.005, and 6.083, respectively.  For UDF PFs i4 and GOODS i6 0, the
emission features are heavily contaminated by sky lines, making it
difficult to ascertain their level of asymmetry and thus whether they
are likely \lya.  We tentatively identify these emission features as
\lya\, placing these objects at \textit{z} = 5.857 and 5.540,
respectively.  However, in the discussion which follows, we shall also
consider the implications that these objects are null-detections or
low-redshift contaminants.

To quantify the asymmetry of the four uncontaminated emission lines,
we adopted the two asymmetry parameters developed by Rhoads et
al. 2003.  The ``wavelength ratio'', $a_{\lambda}$, and ``flux
ratio'', $a_{f}$ are defined as,

\begin{equation}
a_{ w}=\frac{\lambda_{10,r}-\lambda_{p}}{\lambda_{p}-\lambda_{10,b}}
\label{awave}
\end{equation}
\begin{equation}
a_{ f}=\frac{\int_{\lambda_{p}}^{10,r} f_{\lambda}d\lambda}{\int_{\lambda_{10,b}}^{p} f_{\lambda}d\lambda}
\label{aflux}
\end{equation}

respectively.  Here, $\lambda_{p}$ is the wavelength of the peak of
the emission, and $\lambda_{10,b}$ and $\lambda_{10,r}$ are the
wavelengths where the flux exceeds 10$\%$ of its peak value on the
blue and red side of the emission.  The resulting values, shown in
Table \ref{LAE_table}, range from $a_{w}$=1.8-2.6 and $a_{f}$=1.4-2.4.
This is in agreement with values determined from \lya\ emitting
objects at \textit{z} $\approx$ 4.5 \citep{Dawson2004}, and at
\textit{z} $\sim$ 5.7 \citep{Rhoads2003}.  These asymmetry parameters
are inconsistent with those of the \textit{z} $\approx$ 1 O
{\scriptsize II} $\lambda$3727 doublet ($a_{w}\leq1$ and $a_{f}\leq1$;
\citet{Dawson2004}).  Note that for object BD46, the FORS2 spectrum
(see $\S$3.2.1) was used to calculate asymmetry parameters.

The spectrum of UDF PFs i0 was contaminated by bleeding from an
alignment star, impeding detection of \lya\ emission or a continuum.
Thus, this object is excluded from our sample in the analysis below.

\begin{deluxetable*}{cccccc}
\tabletypesize{\scriptsize}
\tablecaption{Objects With Detectable \lya\ Emission\label{LAE_table}}
\tablewidth{0pt}
\tablehead{
\colhead{Object ID} & \colhead{\textit{z}} & \colhead{$f_{Ly \alpha}$\tablenotemark{a,c}} & \colhead{$W_{Ly\alpha}$\tablenotemark{b,c}} &  \colhead{ FWHM\tablenotemark{c}} & $a_{w}$/$a_{f}$\tablenotemark{d}\\
\colhead{} & \colhead{} & \colhead{($10^{-17}$ ergs cm$^{-2}$ s$^{-1}$)} & \colhead{(\AA)} &  \colhead{(\AA)} & 
}
\startdata
BD46        & 5.914 & 0.90 &  24  &  6  & 1.9/1.8\\
BD00        & 5.942 & 3.50 &  150 &  10  &  1.8/1.4\\
GOODS i6 0  & 5.540 & 1.34 &  31 &  19  &  \ldots \\
UDF PFs i4     & 5.857 & 0.70 &  34 &  10 & \ldots  \\
UDF PFs i1     & 6.005 & 1.10 &  65 &  13  & 2.6/2.4\\
UDF PFs i2     & 6.083 & 0.96 &  64 &  10  & 2.3/1.5

\enddata
\tablenotetext{a}{Flux from \lya\ emission.}
\tablenotetext{b}{Measured rest-frame equivalent width.}
\tablenotetext{c}{Not corrected for IGM absorption.}
\tablenotetext{d}{Emission line asymmetry parameters \citep{Rhoads2003}}
\end{deluxetable*}

The two objects UDF PFs i1 ($z_{850} \ = \ 26.9$) and UDF PFs i2
($z_{850} \ = \ 27.1$) are among the faintest \textit{z} $>$ 5 objects
in the $z_{850}$ band to be selected by the Lyman break technique with
confirmed redshifts, very similar to the objects GLARE 3001 and 3011,
with $z_{850}$\ = 26.37 $\pm$ 0.06 and 27.15 $\pm$ 0.12 respectively
\citep{Stanway2004a}.  Furthermore, UDF PFs i2 is faintest object in
our sample (excluding UDF PFs i0), and is the object with the highest
confirmed redshift.

Table \ref{LAE_table} gives the measured redshifts, \lya\ fluxes
($f_{Ly \alpha}$), rest-frame equivalent widths ($W_{Ly\alpha}$), and
FWHMs of the six \lya\ emitting LBGs.  We measured the flux in the
emission lines between the red dotted lines illustrated in Figures 2
and 3.  In some cases, there were residuals from the sky subtraction
that could contaminate the line flux.  We estimated this additional
contribution from the sky lines by measuring the residual flux in
two-dimensional sky-subtracted spectrum outside of the extraction
region used for the one-dimensional spectrum.  All values quoted were
measured directly from the spectra, and do not account for flux lost
due to HI gas absorption in the objects.  Therefore, the quoted fluxes
are lower limits.  The observed emission fluxes from our \lya\
emitting sample are very similar to those found for other \textit{z}
$\sim$ 6 objects selected by the \textit{i}-dropout method
\citep{Bunker2003,Stanway2004b,Stanway2004a}, and for \textit{z}
$\approx$ 5.7 and \textit{z} $\approx$ 6.6 \lya\ emitters selected
using narrowband filters
\citep{Hu2004,Rhoads2003,Lehnert2003,Ajiki2003,Kurk2004,Taniguchi2004,Stern2005,Nagao2004,Rhoads2004a}.
To determine the continuum for each spectrum, we assumed it satisfied
a power law of the form
\begin{equation}
f(\lambda)=f_{o}(\lambda /0.9 \mu m)^{\beta}
\label{power_law}
\end{equation}
with $\beta$ = $-1.1$, which is the average of 198 \lya\ emitting LBGs
at \textit{z} $\sim$ 3 \citep{Shapley2003}.  $f_{o}$ was determined by
fitting the $z_{850}$ magnitude.  The equivalent widths of the \lya\
emissions were calculated directly from these estimated continua.
These widths were not very sensitive to the assumed value of $\beta$.

\subsection{VLT Data}

\subsubsection{BD46}

The two-dimensional FORS2 spectrum of BD46 showed a robust emission
line at the central wavelength of 8406 \AA.  Two- and one-dimensional
extracted spectra for this emission are shown in Figure \ref{5727}a.
A clear asymmetric profile for the line is seen, which we identify as
\lya\, yielding a redshift of \textit{z}=5.914 for this source.
Furthermore, a weak continuum was detected redward of the emission,
with $f=3.4 \ \pm$ 1.0 counts \AA$^{-1}$, shown in Fig. \ref{5727}b.
One item to note is BD46's \lya\ emission was detected with LRIS
before the VLT data was available.  Our methodology accurately
measured this redshift (it was within $\Delta z=0.0005$ of the FORS
value), despite a much weaker signal due to the smaller integration
time in our Keck LRIS data.

\subsubsection{BD03}

While this object was undetected in the LRIS (4 hour integration)
spectrum, the FORS2 spectrum shown in Figure \ref{620} contains a
noisy but clearly flat continuum redward of 8500 \AA, with a sharp
discontinuity at 8485 $\pm$ 3 \AA, where the flux drops to
approximately zero blueward of 8400 \AA.  Using a constant step
function to fit the continuum, we find the average flux density above
the discontinuity is $f^{red}(8500-9100$ \AA) = 5 $\pm$ 1 counts
\AA$^{-1}$.  Below the break, this reduces to $f^{blue}(7700-8300$
 \AA) = 1 $\pm$ 1 counts \AA$^{-1}$, consistent with no detected
flux.

This discontinuity is much larger than would be expected from
UV/optical spectral breaks of galaxies at rest wavelengths 4000 \AA\
[D(4000)], 2900 \AA\ [B(2900)], and 2640 \AA\ [B(2640)]
\citep{Stern2000,Spinrad1997,Spinrad1998}.  Another option is that
this object is a radio-loud broad absorption line quasar.  Such
objects can have large continuum breaks of this amplitude.  However,
such objects also possess very red near-IR colors \citep{Hall1997},
unlike BD03 ($J-Ks$ $<$ -0.03).  Furthermore, the resolved morphology
of BD03 is atypical for luminous AGNs at high redshift, which are
generally unresolved.

The abruptness of the break and the lack of slope in the continuum at
larger wavelengths are strong signs that the discontinuity is due to
absorption by intergalactic HI gas.  Coupling this with the
$i_{775}-z_{850}$ $>$ 2.1 colors (the second reddest in our
spectroscopic sample), we interpret this object as being a starburst
galaxy at \textit{z} = 5.98 $\pm$ 0.1.  Using the LRIS spectrum we
place an upper flux limit of $8.0$ $\times$ $10^{-19}$ ergs cm$^{-2}$
s$^{-1}$ on the \lya\ emission.\footnote{All flux limits are for a 3
$\sigma$ detection extracted over 4 \AA\ ($\approx 3
\Delta\lambda_{res}$).  Note that \citet{Stanway2004b} quoted a 10.5
hr DEIMOS spectra flux limit of $2.0$ $\times$ $10^{-18}$ ergs
cm$^{-2}$ s$^{-1}$, which is for a 5 $\sigma$ detection extracted over
8 \AA.}

\subsubsection{BD38}

This object was also undetected in the LRIS observations, but yields a
strong continuum in the FORS2 spectrum.  Figure \ref{4202} shows the
two- and one- dimensional spectra of the object, along with the ACS
and ISAAC imaging.  A precipitous continuum break is clearly seen at
$\approx$\ 7900 \AA, reducing the continuum from 27 $\pm$ 1 to 0 $\pm$
1 counts \AA$^{-1}$.  Also, several strong absorption features are
found at 8205.77 \AA, 8485.38 \AA, 8498.43 \AA, 8699.72 \AA, 9079.21
\AA, and 9140.84 \AA\ in the spectrum.  Associating the break with the
UV optically thick \lya\ forest, and the absorption lines with the
strong interstellar absorption features typical of LBGs at lower
redshifts \citep[see][]{Shapley2003,Ando2004}, yields a redshift
determination of 5.515 $\pm$ 0.003.

BD38 is an unusually bright \textit{i}-dropout.  Its $z_{850}$-band
magnitude is 24.3, making this object more than 1.3 mag brighter than
any other $i$-dropout in our spectroscopic or photometric sample.  In
fact, to our knowledge, this object is still the brightest \textit{z}
$\sim$ 6 LBG discovered to date with a confirmed redshift.  The
complete spectrum, along with a more thorough analysis of this object
is presented in \citet{DowHygelund2005b}.

\section{Discussion}

\subsection{Confirmation and Contamination}
\label{sec-contamination}

Of the total 21 galaxies in our spectroscopic sample (excluding UDF
PFs i0), we identified spectral features for eight objects.  Of these
eight, we confirmed that six \textit{i}-dropouts from our sample are
\textit{z} $\sim$ 6 galaxies, with the strong possibility that two
more are.  Therefore, the success rate of our survey is assuredly
29$\%$, and most likely 38$\%$.  For the smaller integration times
obtained with LRIS (4 hours), the completeness drops to 25$\%$ (19$\%$
disregarding GOODS i6 0 and UDF PFs i4).  The confirmation rate in our
longer 22.3 hr FORS2 integrations is 100$\%$ (three out of three).
Interestingly, the two reddest $i_{775}-z_{850}$ color objects, the
two brightest $z_{850}$ objects, and the faintest $z_{850}$ object in
our sample are confirmed \textit{z} $\sim$ 6 galaxies.

There was no evidence from the spectra that any of the objects in our
sample are low-redshift contaminants despite the long integration
times.  This is consistent with these objects being \textit{z} $\sim$
6 continuum objects without strong \lya\ emission.  However, if we
associate the emission lines of GOODS i6 0 and UDF PFs i4 with those
of low redshift interlopers (for which we have no evidence) the
contamination of the sample could be as high as 14$\%$, though this is
unlikely.  Support for this point is provided by the low resolution
spectroscopy conducted by \citet{Malhotra2005} using the ACS grism to
test the \textit{i}-dropout selection.  Using a liberal color cut of
$i_{775}-z_{850}>0.9$ (versus our $i_{775}-z_{850}>1.3$ criterion),
they verified that 23 out of 29 candidate $i$-dropouts showed a strong
continuum break indicative of galaxies at \textit{z} $\sim$ 6.  In
addition, they found that a colour cut of $i_{775}-z_{850}>1.3$, while
suffering from modest (20-30$\%$) incompleteness, is subject to a
contamination rate of only 7$\%$ (one out of 14).\footnote{This
contaminant was a star.  The methodology of
\citet{Bouwens2003,Bouwens2004a,Bouwens2005} uses the SExtractor
stellarity parameter \citep{Bertin&Arnouts1996} to exclude stars from
our \textit{i}-dropout selection.  Hence, this contaminant would
likely have been rejected from our spectroscopic sample.  Thus, our
null contamination results are consistent with the findings of
\citet{Malhotra2005}.} Since the estimated contamination in the large
(506 object) $i$-dropout selection of \citet{Bouwens2005} is similarly
small ($\leq$8$\%$), the present findings are perhaps not too
surprising.  This lends strong support to the notion that the objects
in our sample without detectable features are primarily unconfirmed
\textit{z} $\sim$ 6 galaxies, and are not low redshift interlopers.

To see if any of the unconfirmed galaxies in our CL1252 spectroscopic
sample are low-\textit{z} cluster members, we coadded the LRIS spectra
associated with these sources, in an effort to increase the signal to
noise of the observations.  If these objects were predominately
\textit{z} = 1.23 early-type contaminants, a break at $\approx$8800
\AA\ would appear.  No continuum break was detected in the coadded
spectra, nor was any excess continuum found.  Therefore, we concluded
that the majority of the objects in our sample are not cluster
members.  We should note however that this null detection is not very
surprising considering that BD38 and BD03, which are the brightest
objects in our sample and are confirmed \textit{z} $\sim$ 6 objects
via the FORS2 observations, were undetected in the LRIS observations.
The main conclusion of this analysis is that deeper spectroscopy is
necessary to ascertain the nature of these galaxies.

\subsection{Star Formation Rate}

Aside from \lya\ $\lambda$1215.67, no other emission lines were found
in our sample of galaxies with \lya\ emission.  Another sometimes
strong line in our wavelength range is the rest-UV line N {\scriptsize
V} $\lambda1240$.  This emission feature is strong in active galactic
nuclei (AGNs), with line ratios from a composite quasar spectra of
$\langle f_{\rm{Ly\alpha}}/f_{\rm{N V}}=4.0 \rangle$
\citep{Osterbrock1989}.  The flux limits at rest-frame 1240 \AA\ are
$f_{\rm{N V}}<(0.8, 0.8, 1.0, 2.4, 2.2, 1.0) \times\ \rm{10^{-18} \
ergs \ cm^{-2} \ s^{-1}}$, implying lower limits of the
$f_{\rm{Ly\alpha}}/f_{\rm{N V}}>(11.3, 44.8, 13.4, 2.9, 5.0, 9.6)$
line ratios.  Furthermore, no object in our sample was detected in
Chandra or XMM-Newton X-ray imaging data
\citep{Rosati2004,Giaconni2002}.  The lack of any N {\scriptsize V}
$\lambda1240$ emission or X-ray detection suggests that the \lya\
emission is not produced by an AGN.  Thus, we believe the \lya\
photons are primarily from hot young stellar populations consistent
with the \lya\ emitting objects being starbursting galaxies.

Now we estimate the star formation rates of our eight confirmed and 13
unconfirmed objects in our \textit{z} $\sim$ 6 sample.  Two different
methods are used, the first relying the \lya\ emission flux
(\textit{SFR}$_{Ly\alpha}$) to make this estimate, and the second
relying on the $UV$-continuum flux (\textit{SFR}$_{UV}$).  The
relationship between the \lya\ flux and the SFR of a galaxy is given
by
\begin{equation}
SFR_{Ly\alpha} = 9.1 \times \ 10^{-43} \  L_{Ly\alpha} \ M_{\sun} \ \rm{yr^{-1}}
\end{equation}
where $L_{Ly\alpha}$ is the \lya\ luminosity in units of ergs
s$^{-1}$, assuming the Salpeter initial mass function with
($m_{lower}$, $m_{lower}$) = (0.1 $M_{\sun}$, 100 $M_{\sun}$)
\citep{Salpeter1955,Brocklehurst1971,Kennicutt1998}.  The SFRs
determined by this method yield lower limits, due to absorption of
\lya\ photons by dust grains within the galaxy and by the \lya\ forest
\citep{Hu2002}.  The SFRs can also be derived from the UV continuum
luminosities ($L_{UV}$) at $\lambda$ = 1500 \AA, and using the
following relation
\begin{equation}
SFR_{UV} = 1.4 \times \ 10^{-28}\ L_{UV} \ M_{\sun} \ \rm{ yr^{-1}}
\end{equation}
where $L_{UV}$ is in units of ergs s$^{-1}$ Hz$^{-1}$
\citep{Madau1998}.  For the nine objects with ISAAC imaging, the slope
of the continuum was derived from the $z_{850}$, $J$, and $Ks$
magnitudes assuming the UV spectrum can be described by
$f_{\lambda}\propto\lambda^{\beta}$.  The continuum slope used for the
remaining objects was the consistent value of $\beta=-2.0$ found from
\textit{HST}/NICMOS imaging of 26 \textit{i}-dropout objects in the
UDF \citep{Stanway2005,Bouwens2005}.  We adopted a different $\beta$
than that used in \S3.1, because of the observed $\beta$ dependence
on \lya\ emission strength at \textit{z} $\sim$ 3 \citep{Shapley2003}.
For the objects in our sample for which we cannot measure redshifts,
the mean redshift for \textit{i}-dropouts in our selection, \textit{z}
= 5.9 \citep{Bouwens2003}, is assumed.

For objects in the CL1252 field, we used the results of
\citet{Lombardi2005} to correct for gravitational amplification by the
cluster potential.  For each object, we assumed a Non-Singular
Isothermal Sphere lensing model with a free central position.  This
yields a best-fitting velocity dispersion of 1185 km s$^{-1}$.  Again,
for the non-detected objects the mean redshift \textit{z} = 5.9 was
used.  The resulting gravitational magnifications are presented in the
second column of Table \ref{SFR_table}.  Most objects are two or more
Einstein radii away, being only weakly lensed.  However BD22 and BD62
are close to the Einstein radius, and therefore the magnifications
quoted for these sources are highly uncertain.  Hereafter only
delensed values for objects in the CL1252 field are quoted, except for
BD22 and BD62 where observed values are retained.

In Table \ref{SFR_table} we present $L_{Ly\alpha}$, $L_{UV}$,
$SFR_{Ly\alpha}$, and $SFR_{UV}$ for the \lya\ emitting objects in our
sample.  Errors in $L_{UV}$ and $SFR_{UV}$ are derived from the
$z_{850}$-band flux errors.  As is seen elsewhere
\citep{Hu2002,Kodaira2003,Ajiki2003}, for the five \lya\ emitting
objects the ratio $SFR_{Ly\alpha}$/$SFR_{UV}$ ranges widely from
27$\%$ to 127$\%$. This is most likely due to absorption of \lya\
photons by the IGM \citep{Hu2002}.

\begin{deluxetable*}{cccccc}
\tabletypesize{\scriptsize}
\tablecaption{Star Formation Rates for our \textit{z} $\sim$ 6 Sample\label{SFR_table}}
\tablewidth{0pt}
\tablehead{\colhead{Object ID} & \colhead{$\mu$\tablenotemark{a}} & \colhead{$L_{Ly\alpha}$\tablenotemark{b}} & \colhead{$L_{UV}$\tablenotemark{c}} & \colhead{$SFR_{Ly\alpha}$\tablenotemark{d}} & \colhead{ $SFR_{UV}$\tablenotemark{e}}\\
\colhead{} & \colhead{} & \colhead{($10^{42} \  h^{-2}_{0.7} \ \rm{ergs \ s^{-1}}$)} & \colhead{($10^{28} \  h^{-2}_{0.7} \ \rm{ergs \ s^{-1} \ Hz^{-1}}$)} & \colhead{($h^{-2}_{0.7} \ M_{\sun} \ \rm{yr^{-1}}$)} & \colhead{($h^{-2}_{0.7} \ M_{\sun} \ \rm{yr^{-1}}$)}
}
\startdata
BD38        &  1.3 & $<$0.5\tablenotemark{f}& 27.1$\pm$2.7 & $<$0.5  &  38.1$\pm$3.8 \\
BD03        &  1.2  & $<$0.3\tablenotemark{f}   & 10.0$\pm$1.0 & $<$0.4  &  13.4$\pm$1.3 \\
BD00        &  1.1   & 13.24 & 6.8$\pm$0.7 & 12.1 & 9.5$\pm$1.0 \\
BD46        &  1.1   & 3.37 & 6.7$\pm$0.7 & 3.1 & 9.5$\pm$1.0 \\
GOODS i6 0  &\ldots & 4.31 & 6.7$\pm$0.7 & 3.9 & 9.4$\pm$1.0 \\
UDF PFs i4     & \ldots& 2.56 & 6.1$\pm$0.6 & 2.3 & 8.5$\pm$1.0 \\
UDF PFs i1     & \ldots& 4.27 & 3.5$\pm$0.7 & 3.9 & 5.1$\pm$1.0 \\
UDF PFs i2     & \ldots& 3.84 & 3.1$\pm$0.3 & 3.5 & 4.3$\pm$0.4 \\
UDF PFs i0 & \ldots& NS & 12.9$\pm$1.3 &\ldots& 18.0$\pm$1.8 \\
UDF PFs-IDROP1 & \ldots& NS & 12.9$\pm$1.3 &\ldots& 18.0$\pm$1.8 \\
BD27        & 1.5 & NS & 7.6$\pm$0.8 &\ldots& 10.6$\pm$1.1 \\
BD44        & 1.2 & NS & 7.3$\pm$1.5  &\ldots& 10.1$\pm$2.0 \\
BD58        & 1.6 & NS & 6.6$\pm$0.7 &\ldots& 9.3$\pm$1.0 \\
BD66        & 1.2  & NS & 6.2$\pm$0.6 &\ldots& 8.67$\pm$0.9 \\
UDF PFs1 i9    & \ldots& NS & 5.6$\pm$0.6 &\ldots& 7.9$\pm$0.8 \\
BD57        & 1.3 & NS & 5.3$\pm$1.1 &\ldots& 7.7$\pm$1.5 \\
BD22        &   4.22  & NS & 5.1$\pm$1.0\tablenotemark{g} &\ldots& 7.2$\pm$1.4\tablenotemark{g}  \\
BD62        & 35.1  & NS & 4.1$\pm$0.8\tablenotemark{g} &\ldots& 6.2$\pm$1.2\tablenotemark{g} \\
BD40        &   1.6  & NS & 3.9$\pm$0.4 &\ldots& 5.4$\pm$0.5 \\
BD48        &   1.2  & NS & 2.7$\pm$0.5 &\ldots& 3.8$\pm$0.8 \\
BD36        &  1.4 & NS & 2.4$\pm$0.2 &\ldots& 3.4$\pm$0.3
\enddata

\tablecomments{Uncertainties in $L_{UV}$ and $SFR_{UV}$ are derived
from the $z_{850}$-band flux errors.  All quoted luminosities and star
formation rates for CL1252 cluster objects are corrected for possible
lensing, except for BD22 and BD62 where the observed values are used.  An
entry of NS indicates that no spectral features were detected for the
object.  Where no redshift could be determined, the quoted UV
luminosities are for the object at the mean redshift \textit{z} = 5.9
estimated for $i$-dropout selections \citep{Bouwens2003}.}
\tablenotetext{a}{Gravitational magnification due to the CL1252
cluster potential}
\tablenotetext{b}{\rm{\lya\ luminosity.}}
\tablenotetext{c}{UV continuum luminosity at $\lambda$=1500 \AA.}
\tablenotetext{d}{SFR derived from $L_{Ly\alpha}$.}
\tablenotetext{e}{SFR derived from $L_{UV}$.}
\tablenotetext{f}{No detectable \lya\ emission.}
\tablenotetext{g}{Not corrected for gravitational lensing and thus
highly uncertain.}

\end{deluxetable*}

\subsection{Selection Completeness for \lya\ Emitters}
\label{sec-lyacomp}

\begin{figure}
\begin{center}
\figurenum{6}
\includegraphics[width=4in, angle=270]{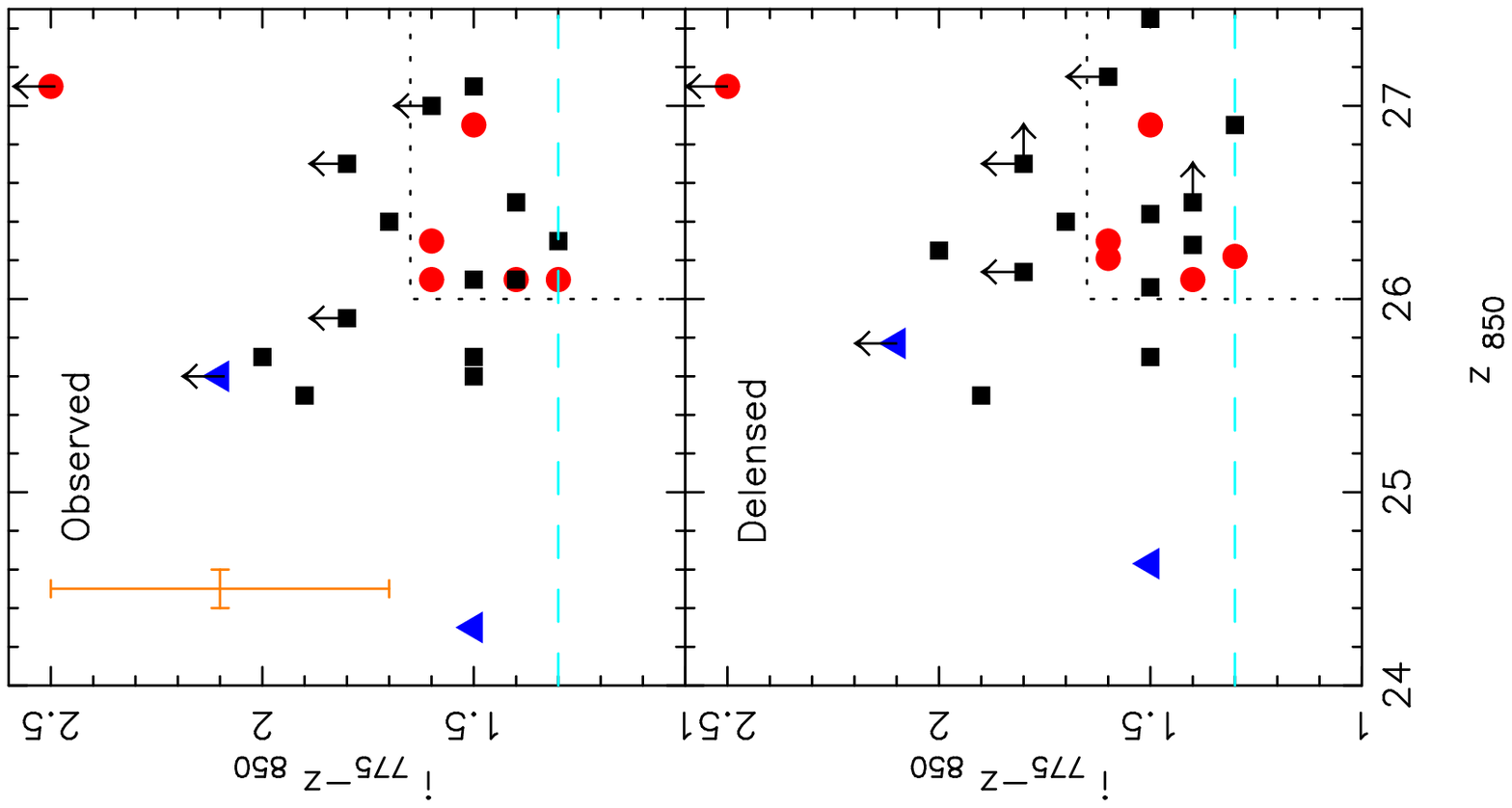}

\caption{\em Observed (upper) and delensed (lower) $i_{775}-z_{850}$
colors versus $z_{850}$-band magnitudes for our spectroscopic sample.
The black squares are sources where no spectral features were found
red circles are the sources with \lya\ emission, and blue triangles
are sources with no detectable \lya\ emission but for which we can
derive a redshift (BD03 and BD38).  Objects undetected in the
$i_{775}$ band are shown at their 2$\sigma$ lower limits and include
upward arrows.  The sources in CL1252 that are not delensed are shown
with right arrows.  The upper left orange error bars are typical for
objects detected in the $i_{775}$ filter.  Our $i_{775}-z_{850}>1.3$
selection criterion is denoted with the blue dashed line.  Aside from
UDF PFs i2, the distribution of \lya\ emitters is constrained to
$z_{850}\geq 26.1$ and $i_{775}-z_{850}\leq 1.6$ (demarcated by dotted
lines).  In the observed plot almost half (9 out of 21) lie outside
this region.  However, this trend is absent after delensing the
objects.  \lya\ emitting objects likely have bluer $i_{775}-z_{850}$
colors than other $i$-dropouts as a result of the contribution of
\lya\ flux to the $i_{775}$-band magnitudes (see Figure
7).\label{color_zmag}}
\end{center}
\end{figure}

\lya\ $\lambda$1215.67 emission is by far the most prominent spectral
feature we can use for redshift determination of \textit{z} $>$ 5
objects.  When the \lya\ emission is strong, it can have an effect on
the observed broadband fluxes.  At \textit{z} $<$ 6, \lya\ flux falls
in the \textit{i}-band, yielding $i-z$ colors that are smaller than
without such emission.  This effect is notable in Figure
\ref{color_zmag}, where we show the $i_{775}-z_{850}$ colors
vs. $z_{850}$-band magnitudes of our sample.  Aside from UDF PFs i2,
which is at \textit{z} = 6.083, the \lya\ emitting objects inhabit the
region $i_{775}-z_{850}\leq 1.6$.  Moreover, from
\citet{Vanzella2006,Vanzella2004,Dickinson2004,Nagao2004,Stanway2004b,Stanway2004a,Bunker2003}
and this work, only two confirmed \lya\ emitting objects have been
detected by the \textit{i}-dropout technique for \textit{z} $<$ 5.78.
Once \lya\ redshifts into the $z_{850}$ band, the decrement can be
larger than it would be without emission, such as for the \textit{z} =
6.33 \lya\ emitting \textit{i}-dropout object SDF J132440.6+273607
discovered by \citet{Nagao2004}.

\begin{figure}
\begin{center}
\figurenum{7}
\includegraphics[width=3.4in, angle=270, scale=1.0]{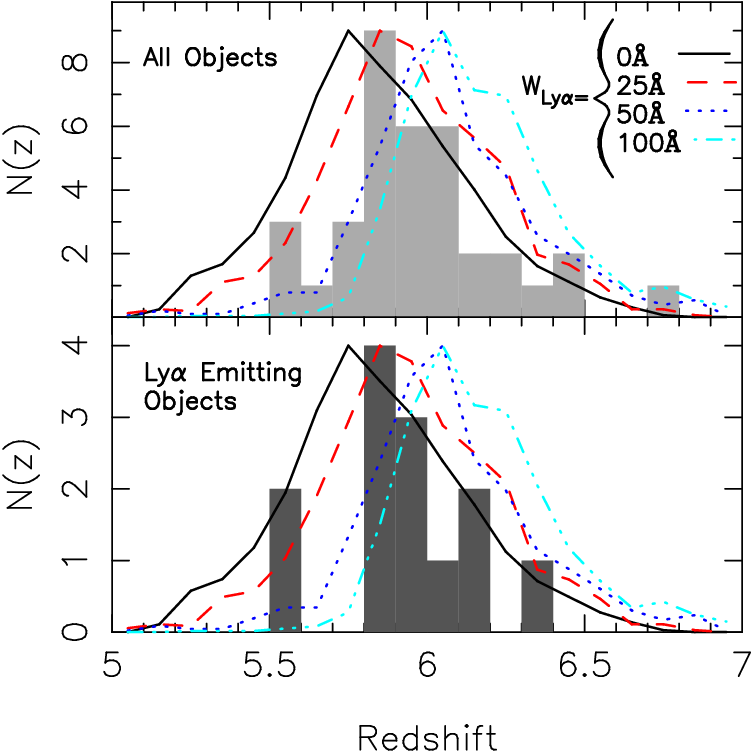}
\caption{\em Observed distribution of spectroscopically confirmed
\textit{z} $\sim$ 6 \textit{i}-dropouts ($i_{775}-z_{850}>1.3$). The
upper histogram contains confirmed sources from this work and
\citet{Stanway2004b}, \citet{Stanway2004a}, \citet{Malhotra2005}, and
\citet{Vanzella2006}.  The lower histogram contains objects with
detectable Ly$\alpha$ emission.  Overplotted are simulated redshift
distributions for four different Ly$\alpha$ equivalent widths
$W_{Ly\alpha}$, indicating the impact of these equivalent widths on
the redshift distribution of the $i$-dropout selection.  Overall, the
simulations respectably encompass the observed distribution, and
suggest that most star-forming galaxies at $z\sim6$ only have modest
Ly$\alpha$ equivalent widths.
\label{dist}}
\end{center}
\end{figure}

In Figure \ref{dist}, we present the redshift distribution of
spectroscopically confirmed \textit{z} $\sim$ 6 \textit{i}-dropouts.
The upper histogram contains objects from this work and
\citet{Stanway2004b,Stanway2004a,Malhotra2005,Vanzella2006}.  The
lower histogram only contains those objects which have measurable
\lya\ emission.  In addition, the expected distribution of
\textit{i}-dropouts from the simulations of
\citet{Bouwens2003,Bouwens2005} are plotted.  From these curves it is
evident that as the strength of \lya\ emission increases, the mean
redshift of the \textit{i}-dropout population increases quite
noticeably, while the width of this distribution narrows somewhat.
Hence, there is a substantial bias against lower redshift (\textit{z}
$\la$ 5.7), strong ($W_{Ly\alpha}\geq50 $\AA) \lya\ emitting
\textit{i}-dropouts being selected as \textit{i}-dropouts.  However,
this effect is small for weaker ($W_{Ly\alpha}=$25 \AA) Ly$\alpha$
emitting objects.  Considering that over half of the \lya\ emitting
objects in our sample have $W_{Ly\alpha}\leq$30 \AA, this should only
result in a modest bias.  In \S~\ref{sec-composition} we discuss this
issue further.

\begin{figure}
\begin{center}
\figurenum{8}
\includegraphics[width=3.4in, angle=270, scale=1]{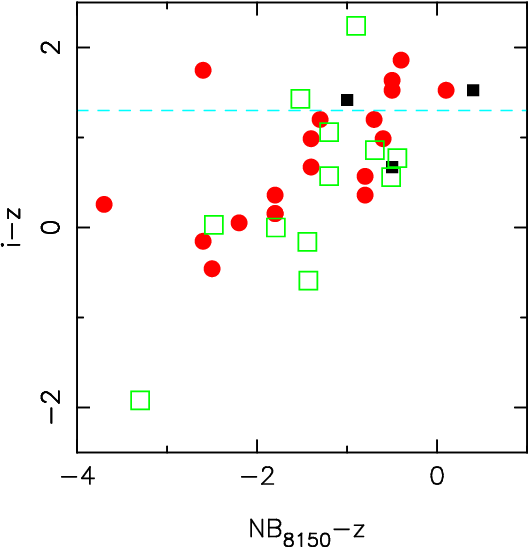}

\caption{\em i-z vs. $NB_{8150}$-z color distribution of two
narrowband selected samples.  The red circles are spectroscopically
confirmed \textit{z} $\approx$ 5.7 \lya\ emitters and the black filled
squares are unconfirmed objects from \citet{Hu2004}.  The green open
squares are narrowband excess candidate \lya\ emitters from
\citet{Ajiki2005}.  No excess flux in the $z_{850}$-band was measured
for two objects in this sample; hence they are excluded from the plot.
A typical \textit{i}-dropout color selection of $i_{775}-z_{850}>1.3$
is shown by the blue dashed line.  Nine out of 33 objects would fall within
our \textit{i}-dropout selection.  Hence, the i-band dropout technique seems to
select $\approx$30$\%$ of the \textit{z} $\approx$ 5.7 \lya\ emitters
found by narrowband surveys.  It is relevant to note that four of
these are the weakest emitters of the confirmed sample of
\citet{Hu2004}.  The other objects do not meet our selection criterion
because of the contribution of \lya\ flux to the \textit{i}-band
magnitude (see Figure \ref{dist}).  As noted in Fig. \ref{dist}, this
selection bias is not a problem for higher redshift (\textit{z}
$\geq$ 6) LAEs.
\label{Hu_comp}}
\end{center}
\end{figure}

To understand how \lya\ emission affects \textit{i}-dropout selection
and how this influence evolves with redshift, we modeled the spectra
of the Hu et al. (2004) (hereafter Hu04) sample of 19 confirmed
\textit{z} $\approx$ 5.7 LAEs, all identified by a narrowband selection.
We plot the colors for Hu04 in Figure \ref{Hu_comp}.
Spectra of these objects were assumed to satisfy the power law given
in Eq. \ref{power_law}.  The flux decrement due to the IGM was modeled
as a simple transmission coefficient, where initial values for the fit
were obtained from the tables of \citet{Songaila2004}.  The continuum
level was set to zero blueward of the Lyman limit $\lambda$912.  The
IGM transmission, $f_{o}$, and total \lya\ emission integrated flux
were varied as three fit parameters until the quoted
\textit{NB$_{8150}$}, Cousins $Z$, and Cousins $I$ magnitudes in Hu04
were reproduced using the IRAF software package Synphot.  These
fiducial spectra were then artificially redshifted or 
blueshifted across the space probed by the
$i_{775}$-dropout technique in intervals of $\Delta$\textit{z} = 0.1.
At each interval, the $i_{775}$ and $z_{850}$ magnitudes were
recomputed using the model for each galaxy.
   
By directly applying our search criterion on these calculated
$i_{775}$ and $z_{850}$ magnitudes, we can determine the efficiency of
the $i_{775}$-drop method at selecting LAEs at various redshifts.
Using the relation

\begin{eqnarray}
n = \int_{z_{1}=5.5}^{z_{2}=6.2} \rho(z) \,\rm{dV}(\textit{z}) &
\approx & 
\sum_{i=1}^7 \rho(\textit{z}_{\rm{i}}) \Delta
\rm{V}(\textit{z}_{\rm{i}}) \nonumber \\ 
& =  & \sum_{i=1}^7 \bar{\rho}*\epsilon(\textit{z}_{\rm{i}}) \Delta \rm{V(\textit{z}_{i})}
\label{expectation_value}
\end{eqnarray}

where \textit{n} is the expected value of LAEs, $\rho(z)$ is the
volume density of Hu04 LAEs at redshift \textit{z}, $\bar{\rho}$ is
the true volume density of Hu04's objects, $\epsilon(z_{\rm{i}})$ is
the fraction of Hu04 LAEs we would select at redshift
\textit{z}$_{\rm{i}}$ (i.e., selection probability), $\Delta
\rm{V}(\textit{z}_{\rm{i}})$ is the volume element for the redshift
interval $z_{\rm{i+1}}-z_{\rm{i}}$, and the sum ranges from
\textit{z}$_{1}$ = 5.5 to \textit{z}$_{7}$ = 6.1.  This relation
yields an expectation value of 3.0 strong LAEs in the CL1252 field.

This expectation value is an upper limit due to two additional
effects.  First, the probability of a \textit{z} $\sim$ 6 object being
selected by the $i_{775}-z_{850}$ dropout technique is a strong
function of redshift, dropping off rapidly at the highest redshifts
probed \citep{Bouwens2003,Bouwens2005}.  This is due to surface
brightness selection effects at work at the high-redshift end of our
\textit{z} $\sim$ 6 $i$-dropout selection \citep[see][]{Bouwens2005}.
Second, \lya\ emission redshifted into the OH bands will be much more
difficult to detect than emission in the region between the bands, and
so the comoving volume probed will be smaller than we assumed.
Mitigating this fact is the quoted flux limit of Hu04 is 2 $\times$
$10^{-17}$ ergs cm$^{-2}$ s$^{-1}$.  This is slightly greater than the
average flux limit for our CL1252 spectra for a \lya\ emission in the
redshift interval probed (1.9 $\times$ $10^{-17}$ ergs cm$^{-2}$
s$^{-1}$).

Hence, we believe the expectation value of 3.0 LAEs to be consistent
with our spectroscopic results of two \lya\ emitting LBGs (BD00 and
BD46) in our CL1252 sample of 12 objects.  Moreover, it appears our
survey is complete in identifying  objects that have
measurable \lya\ emission.  Coupling this with our likely minimal
low-redshift contamination rates (see $\S$\ref{sec-contamination})
suggests that \lya\ emitting LBGs only represent a fraction of
\textit{i}-dropout selected \textit{z} $\sim$ 6 LBGs.

We estimated the total number of expected LAEs over the redshift range
of our \textit{i}-dropout selection assuming the Hu04 volume density
and the \citet{Bouwens2003} covolume.  This yields a strict upper
limit of 5.6 expected LAEs in our CL1252 sample.  From this estimate,
the \textit{i}-dropout selection misses at most $2.6/5.6$, or
$\simeq$46\% of the total LAE population, though most of those missed
are at the low redshift end of the \textit{i}-dropout range.

\section{Trends in \textit{z} $\sim$ 6 Lyman Break Galaxies}

\subsection{Fraction of \lya\ Emitting Galaxies}
\label{sec-composition}
In our sample, we find six objects (four if GOODS i6 0 and UDF PFs i4
are disregarded) with rest-frame equivalent widths strong enough to be
detected as narrowband excess objects ($W_{Ly\alpha}$ $\geq$20 \AA).
Thus 29$\%$ (19$\%$) of our sample (excluding UDF PFs i0) would be
recovered by narrowband surveys.  Similar results of 33$\%$, 33$\%$
and 31$\%$ are found by \citet{Stanway2004b}, \citet{Stanway2004a},
and \citet{Vanzella2006}.  These observations, and the results of the
\lya\ emission completeness test of $\S$\ref{sec-lyacomp}, suggest that
\lya\ emitting objects represent $\approx$30$\%$ of \textit{i}-dropout
spectroscopic samples.

However, just because the fraction of \textit{i}-dropouts with \lya\
emission is 30\% does not imply that the same thing is true for the
$z\sim6$ population as a whole.  In $\S$\ref{sec-lyacomp}, and easily
seen in Fig. \ref{dist}, the $i_{775}-z_{850}>$1.3 color cut biases
our selection against the inclusion of lower redshift (\textit{z} $<$
5.8) objects with strong \lya\ emission ($\geq$50\AA).  This suggests
that the effective selection volumes (and redshift distributions) of
\textit{i}-dropouts with significant \lya\ emission may be quite
different from the distribution without such emission.

This can be accounted for by utilizing the simulations of
\citet{Bouwens2005} (see Fig. \ref{dist}) to determine the effective
search volumes, and corresponding densities, for objects of different
\lya\ equivalent widths.  For the distribution of $W_{Ly\alpha}$ in
our sample (i.e., three objects with $W_{Ly\alpha}$=25 \AA, two with
$W_{Ly\alpha}$=50 \AA, and one with $W_{Ly\alpha}$=100 \AA) these
simulations imply that 32 $\pm$ 10$\%$ of star-forming galaxies at
\textit{z} $\sim$ 6 have $W_{Ly\alpha}$ $\geq$20 \AA, and 7 $\pm$
6$\%$ have $W_{Ly\alpha}$ $\geq$100 \AA.\footnote{We compute the
errors on the fraction of LAEs assuming a binomial distribution.  This
assumes systematic errors are small.}  In the situation where every
\lya\ emitting object in our sample has $W_{Ly\alpha}$=100 \AA\ and
spectroscopy misses every object with $z\geq 6.15$, this fraction
increases to 46 $\pm$ 11$\%$.  This case is an upper limit given our
assumptions.  A lower limit of 25 $\pm$ 10$\%$ is determined, if the
emission line objects GOODS i6 0 and UDF PFs i4 are lower redshift
contaminants.  A similar Ly$\alpha$ emitting fraction of 32 $\pm$
14$\%$ (upper limit of 47 $\pm$ 16$\%$) is found using the results of
\citet{Stanway2004b}.

These results are similar to those obtained from \citet{Shapley2003},
who analyzed spectra of over 1000 \textit{z} $\sim$ 3 LBGs.  Though
\citet{Shapley2003} selected their LBGs photometrically using a
two-color cut, the extra color is used to remove potential
low-redshift contaminants.  As noted in \S~\ref{sec-contamination},
the expected contamination of our survey is small; hence, we believe
our \textit{z} $\sim$ 6 LBG survey is similar enough to warrant
comparison.  \citet{Shapley2003} find that only 25$\%$ of their sample
of \textit{z} $\sim$3 LBGs have $W_{Ly\alpha}$ $\geq$20 \AA, and
$\sim$2$\%$ have $W_{Ly\alpha}$ $\geq$100 \AA.  This is similar to the
value we determine from our observed \lya\ fraction (as well as that
of \citet{Stanway2004b}).  Moreover, the fraction of \textit{z} $\sim$
3 objects with $W_{Ly\alpha}$ $\geq$20 \AA\ is within 2$\sigma$ of our
\textit{z} $\sim$ 6 upper limit.  Therefore, it appears that there is
no strong evolution in the fraction of \lya\ emitting objects between
\textit{z} $\sim$ 3 to \textit{z} $\sim$ 6.

Differing results are found by \citet{Shimasaku2006}, who compare the
\textit{z} $\sim$ 6 LBG luminousity function of \citet{Bouwens2005}
with that derived from their \textit{z} $\approx$ 5.7 LAE sample.
These authors determine that nearly every \textit{z} $\sim$ 6 LBG with
M$_{UV}\la -20$ ($z_{850}$ $\lesssim$ 26.6 mag) should have
$W_{Ly\alpha}$ $\geq$20 \AA.  In fact, \citet{Shimasaku2006} argue
that the fraction of \textit{z} $\sim$ 6 LBGs with $W_{Ly\alpha}$
$\geq$100 \AA\ is $\approx$80$\%$.  Clearly these results are
inconsistent with our analysis above.  At best, the
\citet{Shimasaku2006} fraction is a factor of $\approx$2 too high.  To
be consistent with our survey, the spectroscopic efficiency we
calculated in \S~\ref{sec-lyacomp} must be a factor of $>$10 too low,
a situation we consider highly unlikely given our agreement with the
results of \citet{Hu2004} and \citet{Stanway2004b}.  Hence, we believe
that the Ly$\alpha$ emitting fraction of nearly unity found by
\citet{Shimasaku2006} must be wrong.  This may be at least partially
due to the mild evolution in the luminosity function which will occur
between \textit{z} $\approx$ 5.9 (the mean redshift of the
\citet{Bouwens2005} $i$-dropout selection) to \textit{z} $\approx$
5.7, which is not accounted for by \citet{Shimasaku2006}.

\subsection{Luminosities and Star Formation Rates}

Objects in our sample with measurable \lya\ emission all have observed
$z_{850}$-band magnitudes of 26.1 or fainter, populating the faintest
70$\%$ of the upper panel of Figure \ref{color_zmag}.  Furthermore,
the brightest two objects in our CL1252 sample were confirmed
\textit{z} $\sim$ 6 LBGs without any detectable \lya\ emission.
However, after delensing the sources in our CL1252 sample (lower plot
of Fig. \ref{color_zmag}), this luminosity segregation of \lya\
emitting objects is removed.  The \lya\ emitting LBGs found by
\citet{Stanway2004b} have $z_{850}$-band magnitudes of 25.48 $\pm$
0.03, 26.37 $\pm$ 0.06, and 27.15 $\pm$ 0.12.  Aside from the
brightest object in the \citet{Stanway2004b} sample and the extremely
bright ($z_{850}$ = 24.7) object of \citet{Bunker2003}, all confirmed
\textit{z} $\sim$ 6 \lya\ emitting \textit{i}-dropout objects have
$z_{850}$-band magnitudes in excess of 26.

The average star formation rate derived from the UV continuum
($SFR_{UV}$) of our entire sample (excluding the 6$L^{*}$ object BD38)
is 8.8 $\pm$ 3.9 $h^{-2}_{0.7} \ M_{\sun} \ \rm{yr^{-1}}$.  This value
increases slightly to 9.2 $\pm$ 4.4 $M_{\sun} \ \rm{yr^{-1}}$ if we
exclude \lya\ emitting objects.  The average $SFR_{UV}$ of \lya\
emitting objects is 7.7 $\pm$ 2.2 $M_{\sun} \ \rm{yr^{-1}}$.  Hence,
the $SFR_{UV}$ of \lya\ emitting objects appears to be similar to that
of the \textit{i}-dropout population in general.  \citet{Shapley2003}
find at \textit{z} $\sim$ 3 that \lya\ emitting galaxies with stronger
\lya\ emission have lower star formation rates.  However, as noted by
these authors, this trend could be at least partially due to selection
effects.  Given the relatively small size of our sample, and the large
uncertainties (up to a factor of 10) inherent in UV continuum derived
SFRs \citep[see][]{Papovich2005}, our results do not allow us to make
strong statements about the SFRs of \lya\ emitting LBGs relative to
the population as a whole.

\begin{figure}[ht]
\begin{center}
\figurenum{9}
\includegraphics[width=3.4in, angle=0, scale=1]{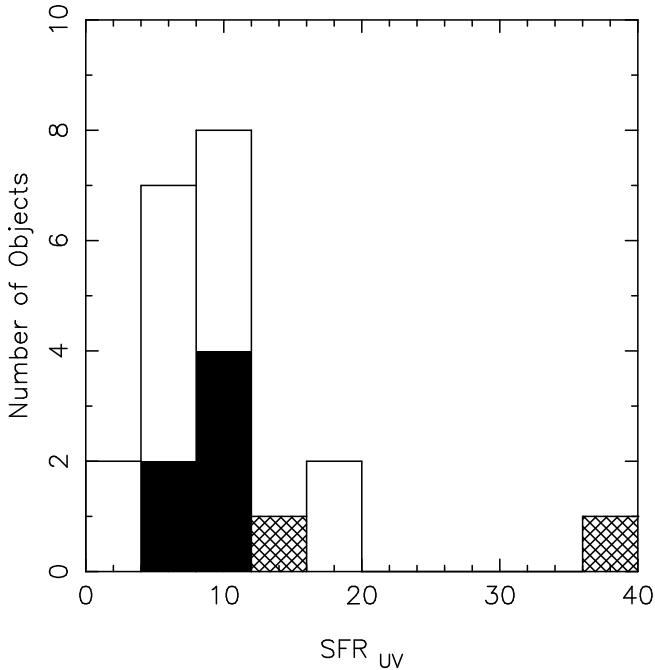}
\caption{\em Histograms of SFR$_{UV}$ for our spectroscopic sample.
Sources in the CL1252 field were corrected for lensing by the cluster
potential, except for objects BD22 and BD62 where the correction was
highly uncertain.  The solid filled histogram are \lya\ emitting
objects, and the cross hatched are the confirmed \textit{z} $\sim$ 6
continuum objects lacking detectable \lya\ emission.  The \lya\
emitting objects appear to have \textit{SFR}$_{UV}$ very similar to
the sample average (excluding the 6$L_{*}$ galaxy BD38).
\label{sfr_ly_hist}}
\end{center}
\end{figure}

\subsection{Correlation between \lya\ Emission and Galaxy Size}

\begin{figure*}
\begin{center}
\figurenum{10}
\includegraphics[width=3.0in, angle=0]{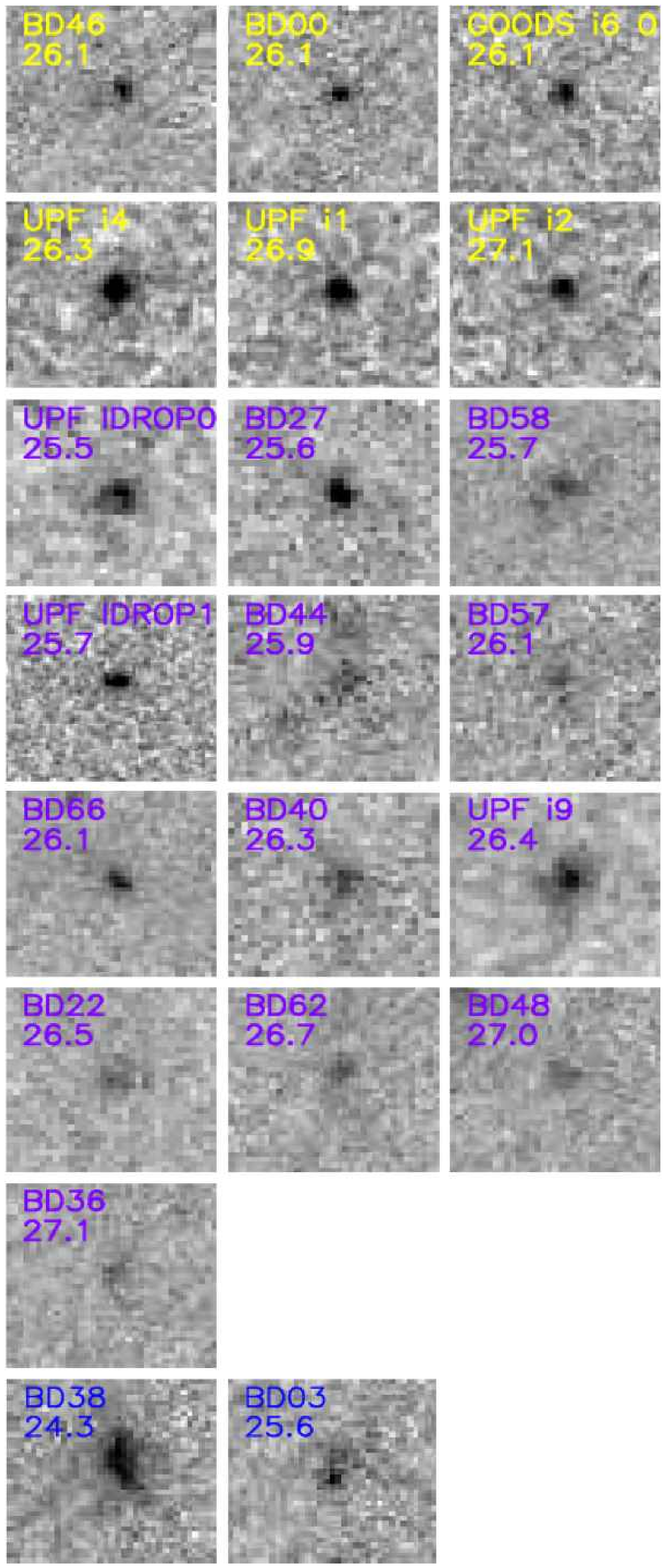}

\caption{\em Thumbnail images of the entire spectroscopic sample in
the $z_{850}$-band (1\farcs5 on a side).  UDF PFs is abbreviated as
UPF.  The upper six thumbnails are our \lya\ emitting sample, the
lowest two are the confirmed \textit{z} $\sim$ 6 objects that are
undetected in \lya\ (BD38 and BD03), and the middle 13 are objects
without any observed spectral features.  The morphologies of BD38 and
BD03 are much more disturbed than the \lya\ emitting LBGs.
Furthermore, the majority of the undetected, fainter objects have a
similar distorted morphology, in contrast to the very compact
appearance of the \lya\ emitting LBGs.
\label{sample_phot}}
\end{center}
\end{figure*}

In Figure \ref{sample_phot} we show $z_{850}$-band images (1\farcs5
thumbnails) for our entire spectroscopic sample (excluding UDF
PFs i0).  The uppermost six thumbnails are the confirmed \lya\
emitting LBGs, the two thumbnails in the bottom row are the confirmed
\textit{z} $\sim$ 6 objects with no detectable \lya\ emission, and the
middle 13 are objects with no clear spectral features.  The \lya\
emitting LBGs all have compact morphologies and show no morphological
disturbances.  However, the morphologies of BD38 and BD03 are very
disturbed and extended, with ``plumes'' extending from a ``core''.
Furthermore, the undetected objects overall have much more extended
morphologies than the objects with detected \lya\ emission.

Similarly, \citet{Stanway2004b,Stanway2004a} confirmed \lya\ emitting
\textit{i}-dropout objects have compact morphologies in the $z_{850}$
band, with 0\farcs09 $\leq$ r$_{hl}$ $\leq$ 0\farcs14.  Moreover,
SBM03$\#$01 also appears compact in \textit{HST} NICMOS 1.1 $\mu$m
F110W-band and 1.6$\mu$ F160W-band imaging \citep{Eyles2005}.
Interestingly, the \lya -emitting \textit{z} = 5.78 galaxy SBM03$\#$03
discovered by \citet{Bunker2003} also has a very compact morphology
(r$_{hl}$ $\leq$ 0\farcs08; marginally resolved in ACS imaging),
despite being exceedingly bright ($z_{850} = 24.7$) for an
$i$-dropout.  This is in contrast to BD38, a similarly bright
($z_{850} = 24.6$) $i$-dropout in our spectroscopic sample, but which
is much more extended (r$_{hl}$ = 0\farcs25) and lacking \lya\
emission.

\begin{figure}
\begin{center}
\figurenum{11}
\includegraphics[width=3.4in, angle=270, scale=1.0]{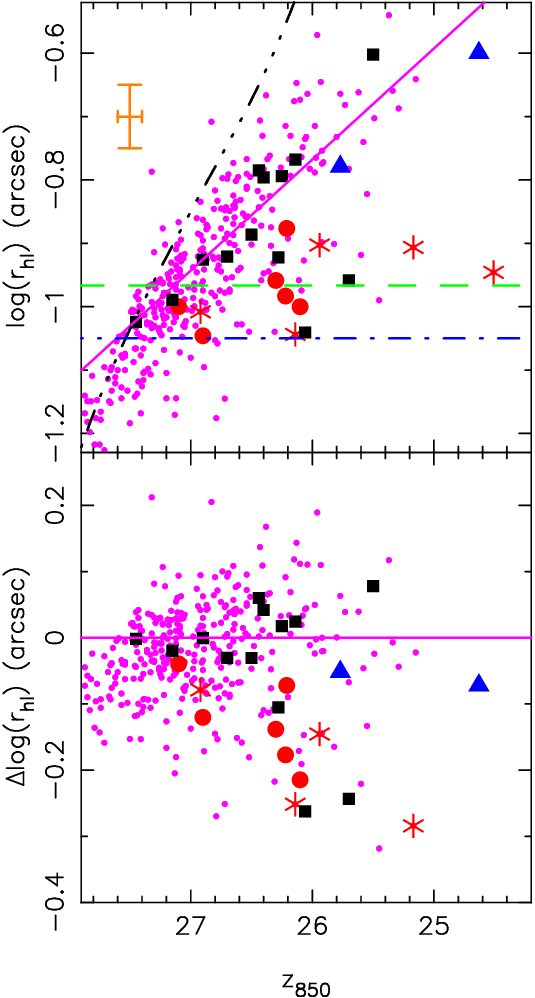}

\caption{\em (Upper panel) Half-light radius versus $z_{850}$-band
magnitude.  The small purple dots represent a sample of 332
\textit{i}-dropouts selected over the GOODS fields
\citep{Bouwens2005}.  The red stars are previously
spectroscopically-confirmed \textit{z} $\sim$ 6 \lya\ emitting LBGs
\citep{Bunker2003,Stanway2004b,Stanway2004a,Vanzella2006}.  The other
symbols are the same as Figure \ref{color_zmag}.  The black
dot-dash-dot-dot line is the surface brightness selection limit of the
\citet{Bouwens2005} sample.  The purple line denotes the best linear
fit to this sample (purple dots; other objects excluded).  The
vertical orange error bar of $\sigma$ = 0.05 dex is the scatter in the
fit. The horizontal error bar is the typical photometric uncertainty.
The blue dash-dot line is the FWHM of the $z_{850}$-band PSF and
represents a reasonable floor below which our size measurements become
quite uncertain.  (Lower panel) Residuals from the linear fit.  The
\citet{Bunker2003} object is outside the plot window ($\Delta \log
{\rm r}_{hl}$ = $-$0.49).  All Ly$\alpha$ emitting objects have
half-light radii smaller than is typical for the 332
\textit{i}-dropouts of \citet{Bouwens2005}.  A Rank-Sum test performed
on the objects from our spectroscopic sample supports the hypothesis
(at $\geq$97\% confidence) that Ly$\alpha$ emitting
\textit{i}-dropouts (6 objects from our sample) are smaller than the
objects without such emission (14 objects from our spectroscopic
sample).  Furthermore, the \lya\ emitting objects have a fairly
constant size (r$_{hl}$ = 0.11 $\pm$ 0.01) over the entire range of
luminosities probed.
\label{rhl_zmag}}    
\end{center}
\end{figure}

In the upper panel of Figure \ref{rhl_zmag} we plot the r$_{hl}$
versus $z_{850}$-band magnitude for sources in our spectroscopic
sample.  The red circles are \lya\ emitting LBGs, the blue triangles
represent BD38 and BD03 (sources with a continuum break), and the
black squares are \textit{i}-dropouts in our sample without any clear
spectral features.  The CL1252 objects have been delensed.  Shown with
purple dots are the 332 \textit{i}-dropouts identified by
\citet{Bouwens2005} over the two GOODS fields.  Also included
(\textit{red stars}) are the four previously confirmed \lya\ emitting
LBGs of \citet{Bunker2003} and \citet{Stanway2004b,Stanway2004a} noted
above, where the half-light radii and $z_{850}$-band magnitudes quoted
are taken from \citet{Bouwens2005}\footnote{Object GOODS-N i\arcmin
Drop 6 of \citet{Stanway2004b} is not included, due to its uncertain
line identification}.  The purple line in Fig. \ref{rhl_zmag} denotes
a linear fit to the $i$-dropout sample (purple dots) of
\citet{Bouwens2005}.  Objects with half-light radii smaller than 
the FWHM of the $z_{850}$-band PSF (0\farcs09; blue dash-dot line
in figure), fainter than our magnitude limit ($z_{850}$ $>$ 27.8), or
brighter than $z_{850}=25.0$ (due to a scarcity of bright sources)
were not included in the fit.  No sources from our spectroscopic sample
were included in the fit.

We want to test the hypothesis that sources with \lya\ emission are
smaller on average than the $i$-dropout population in general.  We chose
not to use a $\chi^2$ test, because it assumes a Gaussian scatter
around the best-fit distribution.  A better test is the Rank-Sum test
utilizing only the half-light radii of our spectroscopic sample.
Using this test to compare our spectroscopic sample with the
\citet{Bouwens2005} $i$-dropout sample, we find that the half-light
radii of the galaxies in our spectroscopic sample are significantly
(at $\geq 94$\% confidence) smaller than those of \citet{Bouwens2005}.
Restricting the Rank-Sum test to include only objects in our
spectroscopic sample also supports the hypothesis that \lya\ emitting
\textit{i}-dropouts are smaller than the rest of our sample at
$\geq$97$\%$ confidence.  Hence, \lya\ emitting \textit{i}-dropouts
seem to be morphologically distinct from other \textit{i}-dropouts.

One potential exception to this rule is the object UDF 5225 of
\citet{Rhoads2005}, which has a ``plume'' of 1\arcsec\ extending from
a compact ``core''.  This object was selected as a $V_{606}$-dropout,
has a spectroscopic redshift of \textit{z} = 5.480, and is too blue
($i_{775}-z_{850} \approx 0.5$) to be selected by \textit{i}-dropout
methods.  However, even if we assumed that this object passed our
color selection, the surface brightness of the plume is below our
limiting surface brightness, due to shallower photometry than the UDF.
Our deepest photometry (UDF PFs) of the object would only contain the
brighter compact ``core''.  Hence, this object would not appear
abnormal for \lya\ emitting objects in our sample.

Note that the nature of the vast majority of the objects (332 in
total) used in our larger $i$-dropout sample is not known; however
from the discussion above, approximately 30$\%$ of these objects
should be \lya\ emitting LBGs.  If so, this would mitigate the
differences observed between the sizes of \lya\ emitting LBGs and
those from the $i$-dropout population in general (which include a
significant fraction of \lya\ emitting sources).  Hence, the
size-luminosity discrepancy between \lya\ emitting and non-emitting
LBGs would be larger than quoted here.

This morphological difference may be due to the influence of the \lya\
emission on the $z_{850}$-band flux.  To date, the only confirmed
\textit{z} $\sim$ 6 \lya\ emitting \textit{i}-dropout with NICMOS
imaging is SBM03$\#$01 \citep{Eyles2005}, which, as noted above, is
compact in both 1.1 $\mu$m F110W-band and 1.6$\mu$m F160W-band
imaging.  Hence, the compact morphology of this LBG cannot simply be
the result of its \lya\ emission alone.  NICMOS imaging of many more
\textit{i}-dropouts with confirmed \lya\ emission is needed to
determine the overall relevance of \lya\ emission to the
$z_{850}$-band morphology of these objects.

One potential physical explanation of this size deviation could lie in
the masses of these objects.  \lya\ emitting objects at \textit{z}
$\sim$ 3-4 are found to have smaller stellar masses than objects
lacking this emission \citep{Overzier2006,Gawiser2006}.  This suggests
that these objects may be associated with less massive dark matter
haloes.  One would make a similar inference for the masses of these
objects using the observation that these objects are less dusty on
average \citep{Shapley2003,Gawiser2006}, thereby having fainter
dust-corrected luminosities, than sources with no observable \lya\
emission.  Assuming this translates to \textit{z} $\sim$ 6, and using
the well-known correlation between size and mass \citep{Mo1998}, we
would expect \lya\ emitting objects to appear smaller on average,
which is what we observe.

We note that the recent results of \citet{Lai2006} on a sample of 12
$z\sim5.7$ LAEs over the HDF-North GOODS area could be seen to support
this finding.  \citet{Lai2006} found that 3 of the 12 sources were
clearly detected in the IRAC data and inferred masses of $10^{9}$
$M_{\odot}$ to $10^{10}$ $M_{\odot}$ through detailed stellar
population modelling.  While these masses are comparable to similar
luminosity $i$-dropouts studied over the GOODS fields
\citep{Yan,Eyles2006}, the vast majority of LAEs in the
\citet{Lai2006} sample (9 out of 12) are not detected in the GOODS
IRAC imaging and hence will be significantly less massive than the 3
IRAC-detected LAEs.  Though clearly a more careful comparison between
these populations is needed, this suggests that $z\sim5.7$ LAEs, on
average, are somewhat less massive than the typical $z\sim6$
star-forming galaxy.

\subsection{Preselecting Galaxies with \lya\ Emission}

One interesting aspect of this morphology dependence on \lya\ emission
strength is in its potential to preselect \lya\ emitting
\textit{i}-dropouts for spectroscopic follow-up, similar to what is
achieved in narrowband surveys.  Confirmation rates of
\textit{i}-dropout selected \textit{z} $\sim$ 6 LBGs are rather low
($\approx$30$\%$) due to the low S/N of these objects in the continuum
and the small fraction of LBGs with \lya\ emission (\S5.1).  From our
spectroscopic sample, we note that a simple size cut $\Delta
\log(\rm{r}_{hl})$ $\leq$ -0.05 and magnitude cut $z_{850}$ $\geq$
25.9 would select five (83$\%$) of the \lya\ emitting LBGs, while
rejecting all but two (15$\%$) of the sources without detectable \lya\
emission.  This would increase our success rate for spectroscopic
confirmation by more than a factor of three.

\section{Conclusions}

We have obtained spectroscopic observations of 22 \textit{i}-dropouts
drawn from two deep ACS fields.  These dropouts were selected to have
$i_{775}-z_{850}$ colors greater than 1.3.  Spectroscopic redshifts
for eight \textit{z} $\sim$ 6 objects in the RDCS 1252-2927 and Ultra
Deep Field Parallel fields were derived.  We thereby confirm the
effectiveness of the \textit{i}-dropout technique of
\citet{Bouwens2003,Bouwens2004a} in isolating a statistically relevant
sample of 5.5 $\leq\textit{z}\leq$ 6.2 star-forming galaxies.

No clear case of contamination by low-redshift sources was found in
this sample.  Together with the results of the complementary ACS Grism
survey of \citet{Malhotra2005}, this suggests our spectroscopic sample
is dominated ($\geq$90$\%$) by galaxies at \textit{z} $\sim$ 6.  Six
of the confirmed \textit{z} $\sim$ 6 objects possess measurable \lya\
emission, with $z_{850}$-band magnitudes ranging from 26.1 to 27.1.
The two brightest objects in our CL1252 sample are continuum sources,
lacking detectable \lya\ emission, but show clear evidence for a Lyman
break.

We compare our findings with other \textit{z} $\sim$ 6
\textit{i}-dropout surveys, \textit{z} $\approx$ 5.7 narrowband
surveys, and with the global properties of LBGs at \textit{z} $\sim$ 3
to determine the nature and composition of \textit{z} $\sim$ 6
galaxies.  Our findings are as follows:

\begin{enumerate}
\item Significant ($W_{Ly\alpha}\gtrsim20$\AA) \lya\ emission is
  detected in the spectra of only 30$\%$ of \textit{i}-dropout
  objects.  Utilizing the model redshift distributions of
  \citet{Bouwens2005} to control for selection biases, we infer that
  only 32$\pm$10\% of star-forming galaxies at $z\sim6$ show
  significant ($W_{Ly\alpha}\gtrsim20$\AA) \lya\ emission (\S5.1:
  Figure~\ref{dist}), with an upper bound of 46$\pm$11\%.  Moreover,
  the $W_{Ly\alpha}$ of these objects on average are much smaller than
  those found for narrowband-selected LAEs.  Since these trends are
  also evident in the LBG population at \textit{z} $\sim$ 3
  \citep{Shapley2003}, this suggests that there is no strong change in
  the fraction of \lya\ emitting objects in LBG population from
  \textit{z} $>$ 5 to \textit{z} $<$ 4.

\item The \textit{i}-dropout technique misses $\approx$70$\%$ of
  narrowband selected \textit{z} $\approx$ 5.7 LAEs.  Moreover, the
  LAEs \textit{i}-dropout surveys do select show the weakest \lya\
  emission.  This is a consequence of the \lya\ emission in the
  $i_{775}$-band flux at $z\le5.9$.  However, the selection efficiency
  increases strongly with redshift, due to \lya\ emission shifting
  from the $i_{775}$ to $z_{850}$ bands.  By \textit{z} $\sim$ 6,
  every $z_{850}$-band detected \lya\ emitter would be sampled by
  \textit{i}-dropout techniques.  Using the simulations in \S
  \ref{sec-composition}, we expect to miss 10-46\% of
  the LAE population with our \textit{i}-dropout selection.

\item \lya\ emitting LBGs have similar $z_{850}$-band magnitudes and
UV SFRs to those without detectable emission.  However, no \lya\
emitting object in our sample has $z_{850}$ $<$ 26.1.

\item The size of \lya\ emitting objects is more compact than
predicted from the observed size-luminosity relation of 332
\textit{i}-dropout galaxies.  One possible explanation for this trend
is that sources with \lya\ emission may be systematically less massive
than the typical \textit{i}-dropout.

\end{enumerate}

Most importantly, these results suggest that \lya\ emitting objects
only constitute a modest fraction of the LBGs at \textit{z} $\sim$ 6
and thus do not provide us with a representative sample.

These results have important implications for how surveys for
star-forming galaxies at $z\sim6$ should be conducted.  First, it
appears that our spectroscopic success rates can be significantly
enhanced by targetting $i$-dropouts with compact $z_{850}$-band
morphologies.  From our sample, we find the criteria $z_{850}$ $\geq$
25.9 and $\Delta \log {\rm r}_{hl}$ $\leq$ -0.05 selects $\geq$80$\%$
of \lya\ emitting objects, while rejecting $\geq$80$\%$ of the
remaining sample.

Second, it appears wide-area narrowband surveys miss $\approx$70$\%$
of the broadband-selected LBG population at \textit{z} $\sim$ 6, and
as seen in Figure \ref{rhl_zmag} the objects they do find are
intrinsically different from the \textit{i}-dropout LBG population as
a whole.  On the other hand, broadband surveys miss the strongest
\lya\ emitters at 5.5 $\leq$ \textit{z} $\leq$ 5.8 that narrowband
surveys sample.  Therefore, it appears a combination of surveys --
taking advantage of the strengths of both narrowband and broadband
selection techniques -- will be necessary to obtain a complete
characterization of the star-forming population at $z\sim6$.
 
\acknowledgements

We would like to thank Dan Kelson for useful discussions concerning
spectral reduction, and David Koo for his helpful insights.  ACS was
developed under NASA contract NAS5-32864, and this research was
supported by NASA grant NAG5-7697.  The authors wish to recognize and
acknowledge the very significant cultural role and reverence that the
summit of Mauna Kea has always had within the indigenous Hawaiian
community.  We are most fortunate to have the opportunity to conduct
observations from this mountain.

\clearpage

\begin{figure}
\begin{center}
\figurenum{2} \includegraphics[width=6in, 
scale=0.8]{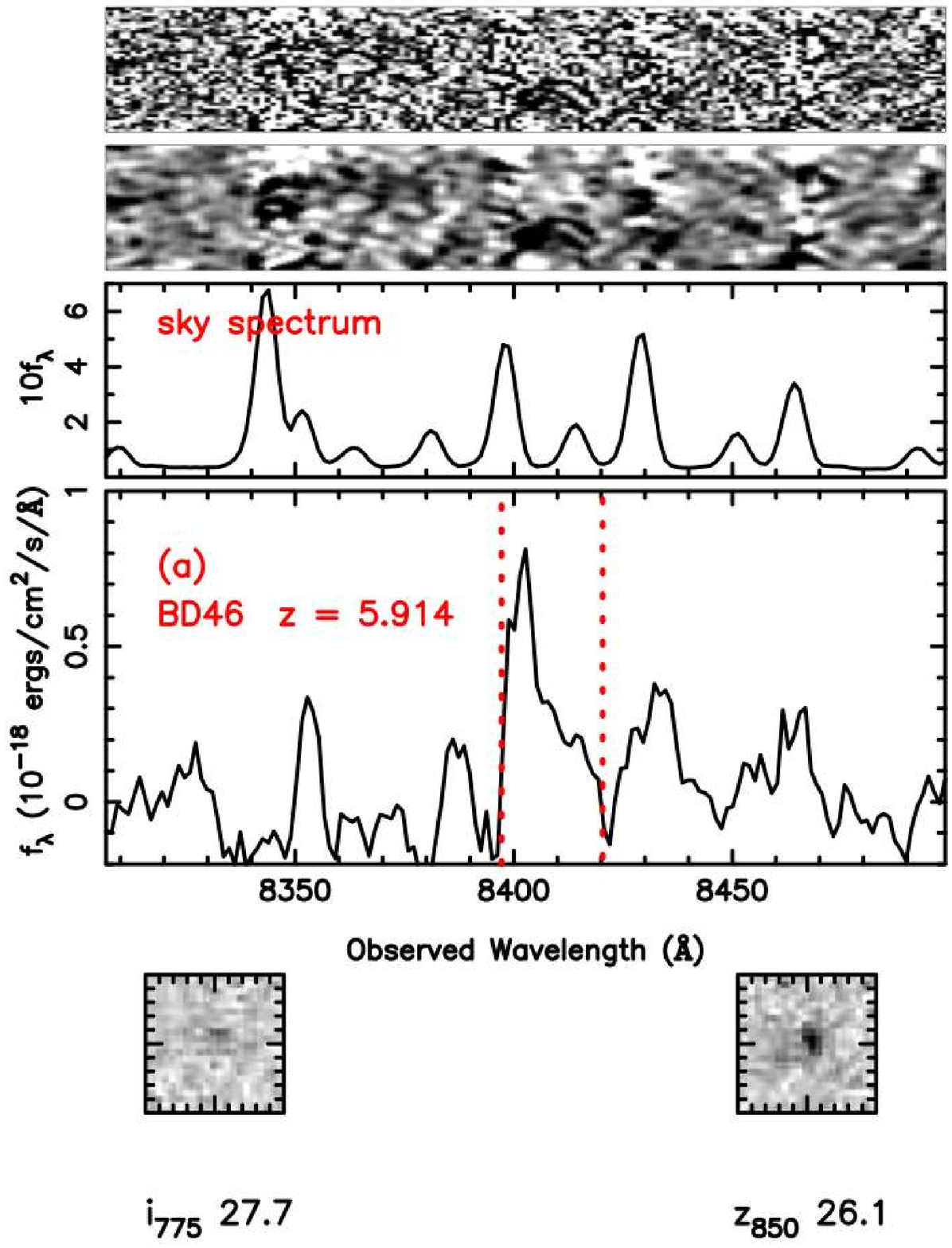}
\caption{\em LRIS spectra and ACS imaging of emission line objects:
two-dimensional unsmoothed spectra (upper panel), candidate \lya\
emission (lower panel), and ACS imaging (lower postage stamps).  Also
shown is the two-dimensional $\sigma$=1 pixel Gaussian smoothed
spectra of the sky (middle panel).  Objects are (a) BD46, (b) BD00,
(c) GOODS i6 0, (d) UDF PFs i4, (e) UDF PFs i1, and (f) UDF PFs i2.
The postage stamps are $\rm{1\farcs0 \times 1\farcs0}$.  All upper
limits given for the $B_{453}$, $V_{606}$, and $i_{775}$ band fluxes
are 2 $\sigma$.  The exposure times for the objects were 7200 s,
except for the two CL1252 objects (BD46 and BD00) where the exposure
times were 16200 s.  Each one-dimensional spectrum has been smoothed
with a 3 pixel boxcar filter, except BD00 and BD46.  Vertical dotted
lines delineate the region used for measuring line fluxes, equivalent
widths, and FWHMs.  For the UDF PFs objects, the blue cross hatched
rectangles represent regions where strong skylines are present.  The
$i_{775}-z_{850}$ colors and the asymmetry of these line profiles are
consistent with the emission lines from BD46, BD00, UDF PFs i1, and
UDF PFs i2 being \lya\ $\lambda$1215.67. The other two emission lines
(GOODS i6 0 and UDF PFs i4) are difficult to identify, due to skyline
interference.  We consider the implication that the features in these
latter two sources are spurious and \lya\ is undetected.
\label{LAE}}

\end{center}
\end{figure}
%\begin{figure}
\clearpage
\begin{center}
\includegraphics[width=6in,scale=0.8]{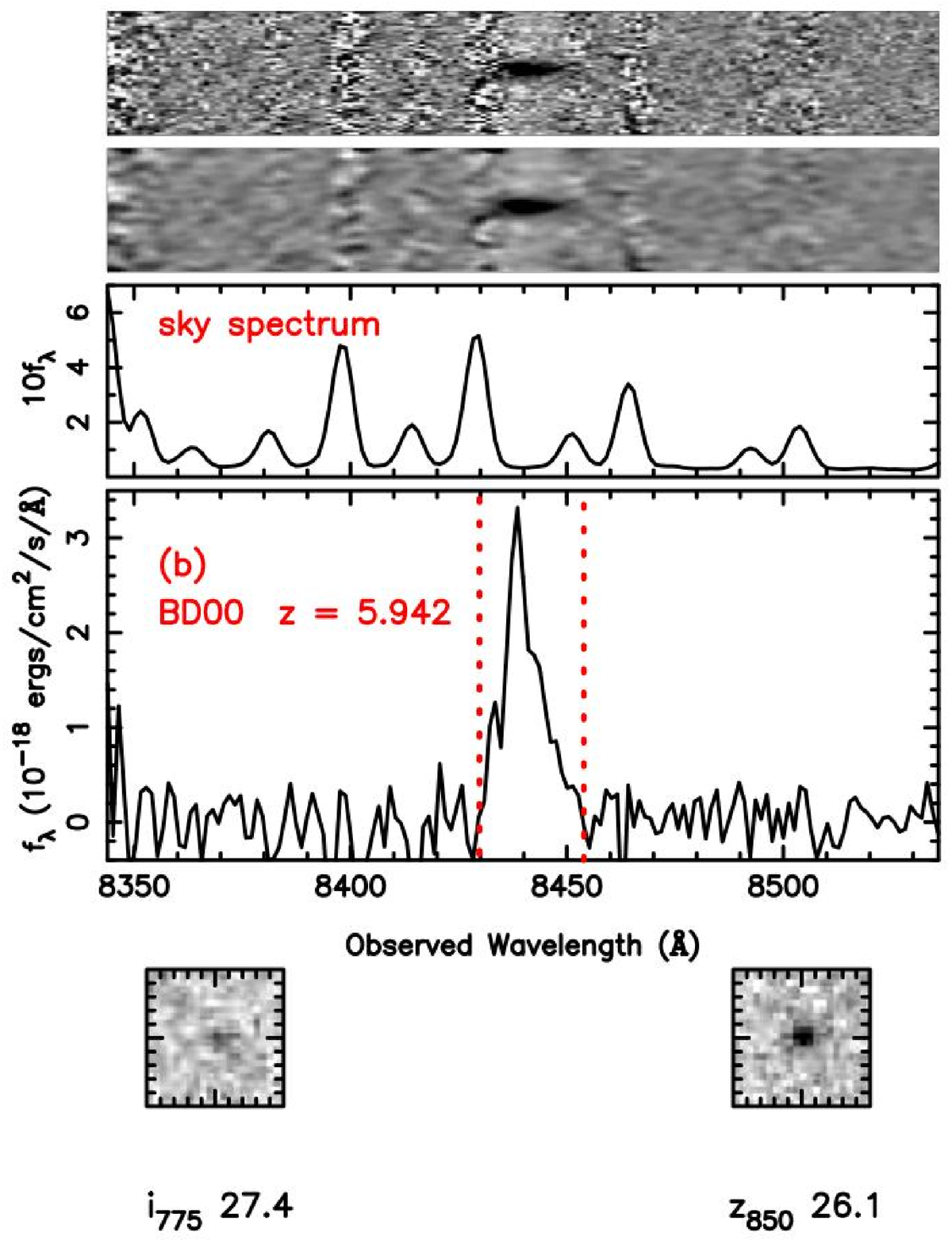}

\includegraphics[width=6in,scale=0.8]{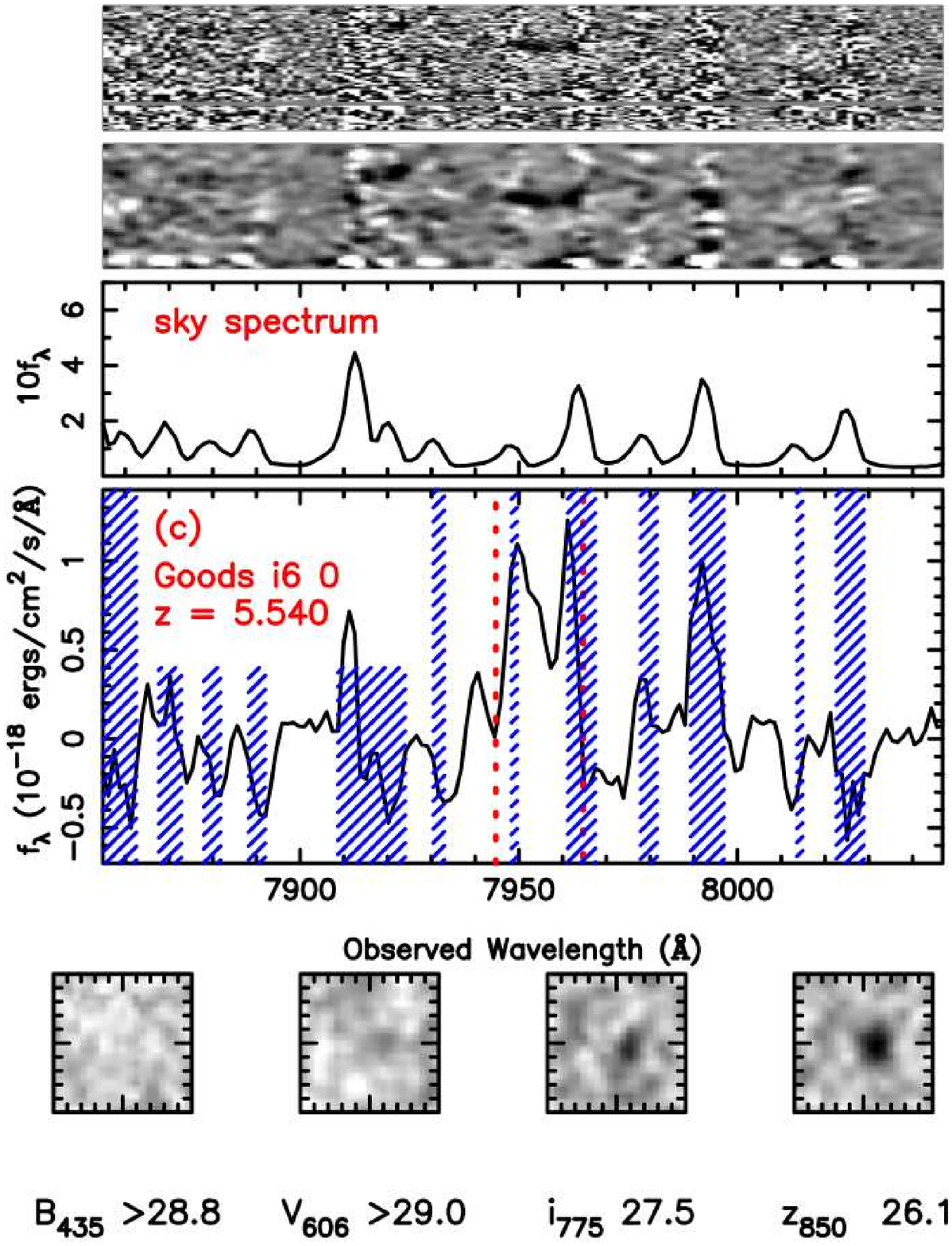}

\includegraphics[width=6in,scale=0.8]{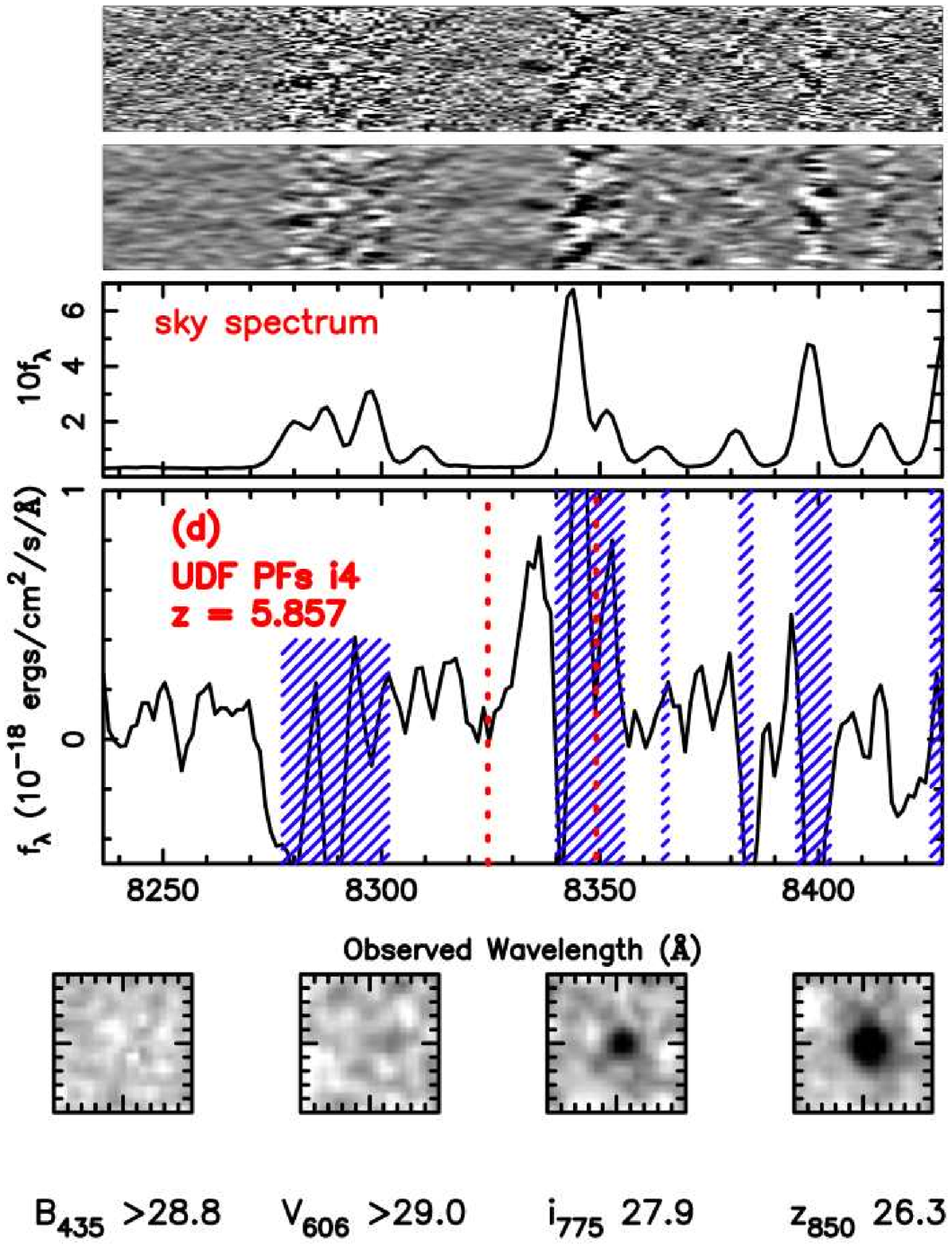}

\includegraphics[width=6in,scale=0.8]{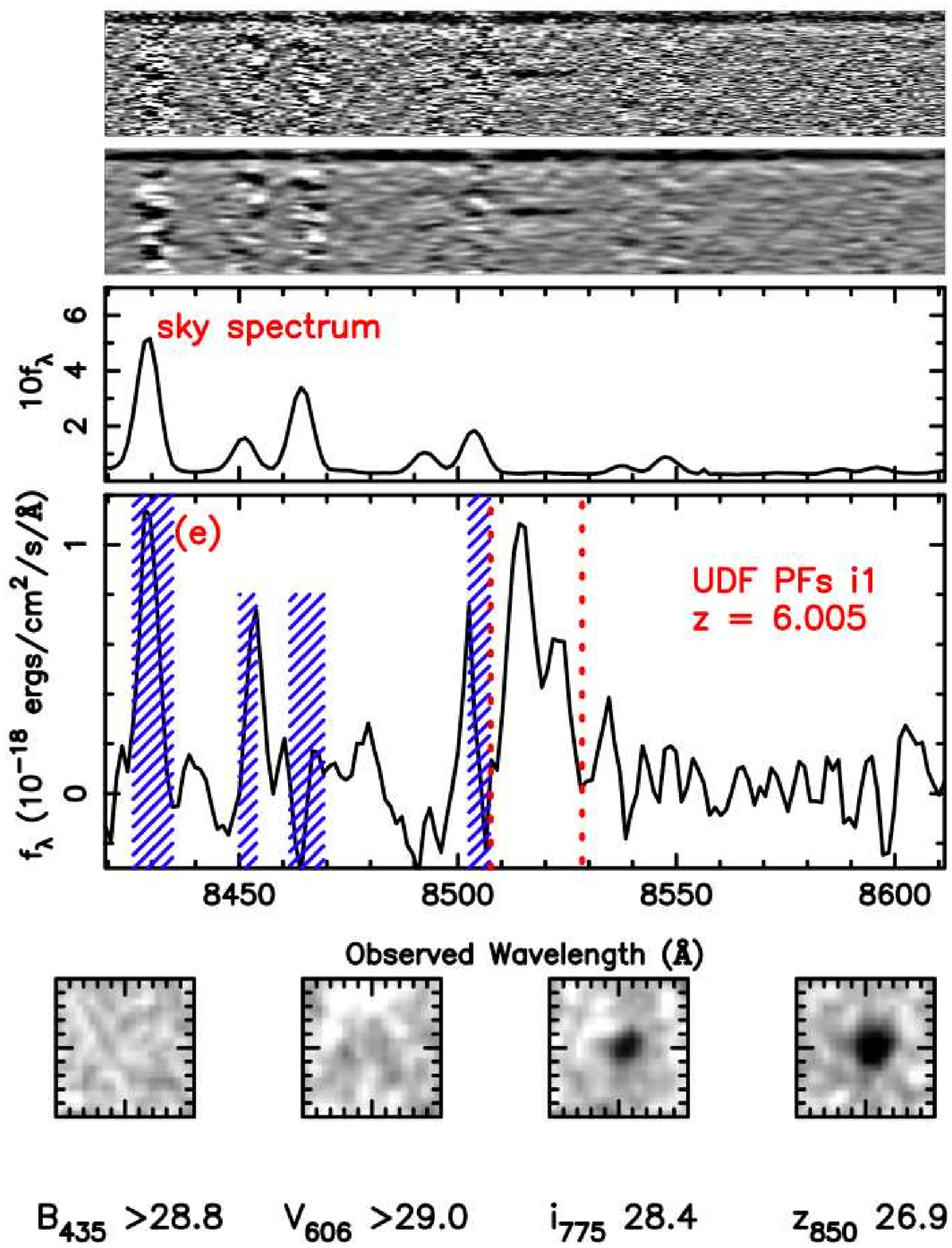}

\includegraphics[width=6in,scale=0.8]{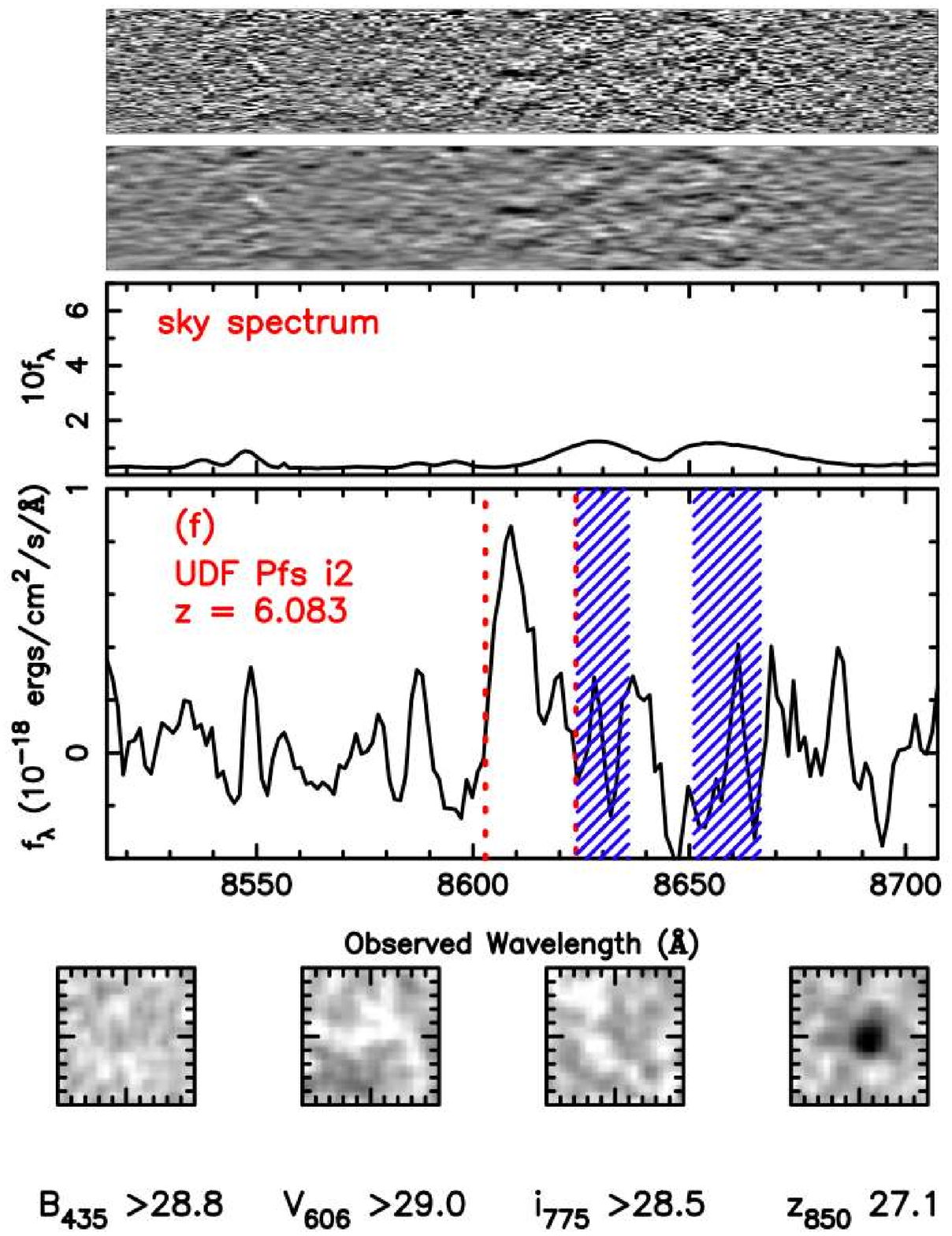}
\end{center}
%\end{figure}
\clearpage

\begin{figure}
\begin{center}
\figurenum{3}
\includegraphics[width=6.0in, scale=1.0]{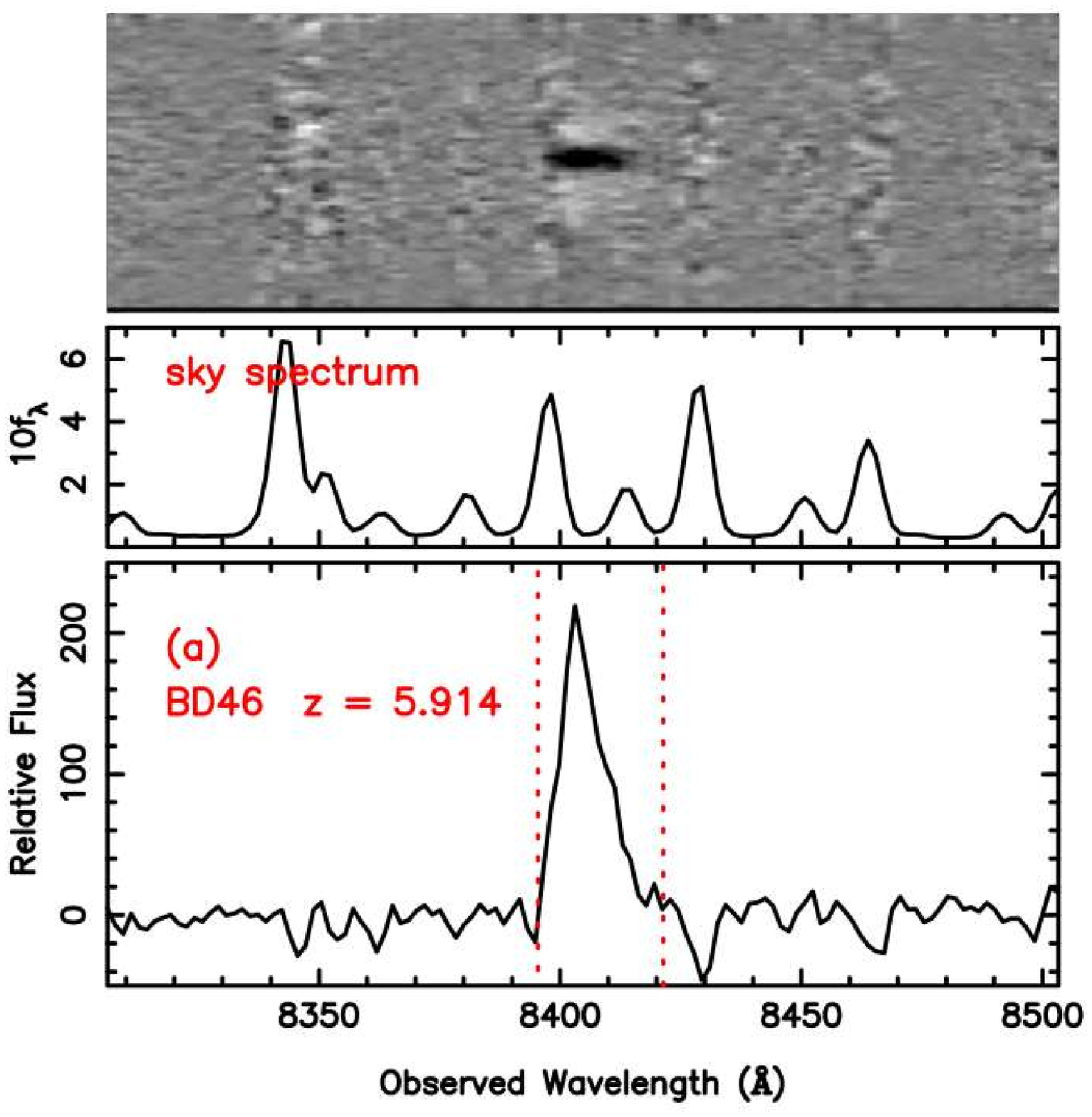}

\caption{\em (a) FORS2 two- and one dimensional spectra of BD46 (LRIS
spectrum shown in Fig. \ref{LAE}).  The total exposure time is 22.3
hours.  The \lya\ emission occurs at 8406\AA, consistent with the LRIS
spectrum.  (b) Shows the faint continuum of BD46 redward of the
emission.  The two dimensional spectrum has been smoothed with a
$\sigma$=1 gaussian.  The one dimensional spectrum has been smoothed
with a 10-pixel boxcar filter.  The red dashed line delineates the
continuum of $f=3.4 \pm 1.0 \ counts \ $\AA$^{-1}$.
\label{5727}}

\end{center}
\end{figure}

\clearpage
\begin{center}
\includegraphics[width=6.0in,  scale=1.0]{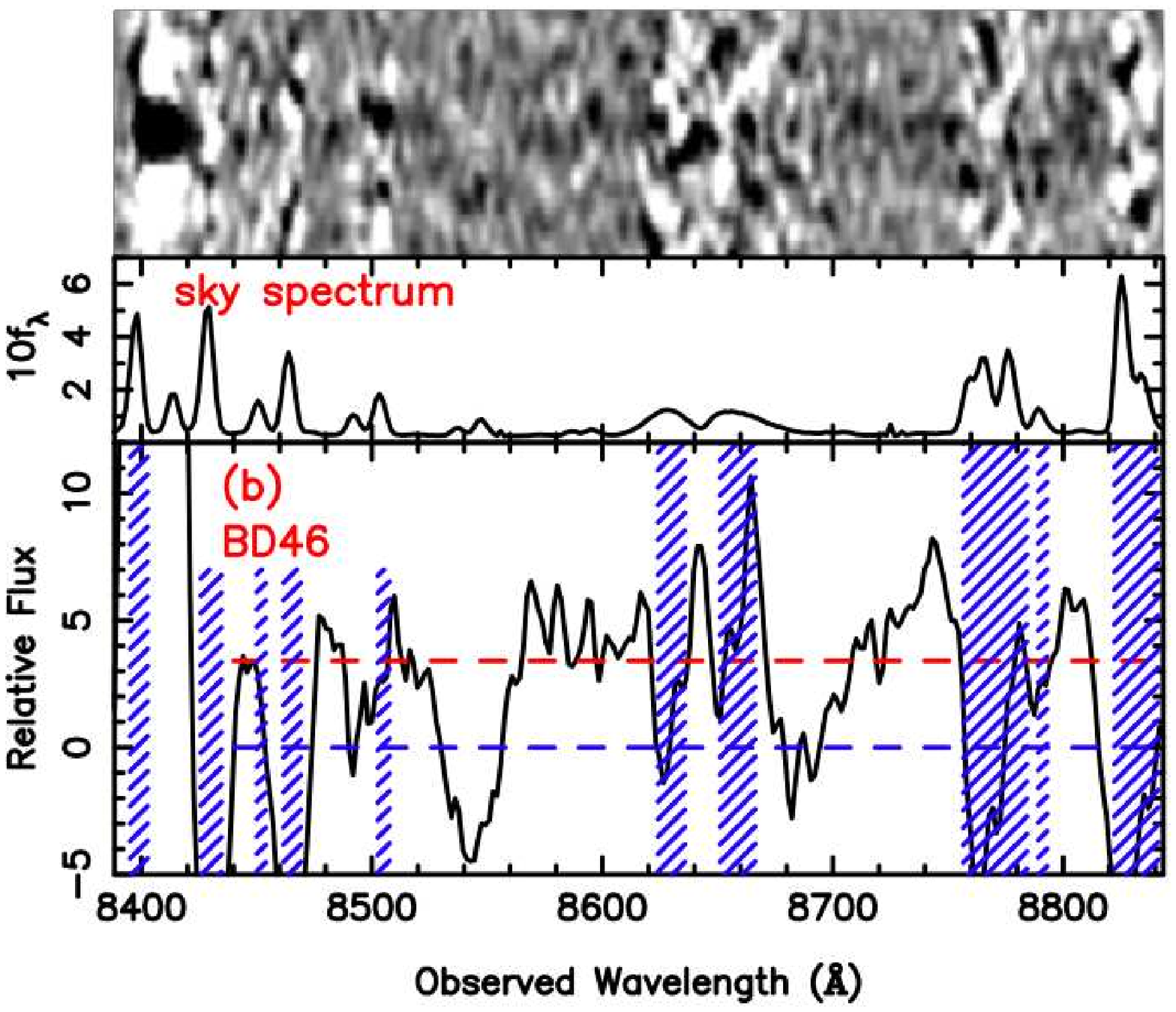}
\end{center}

\begin{figure}
\begin{center}
\figurenum{4}
\includegraphics[width=6in, scale=0.8]{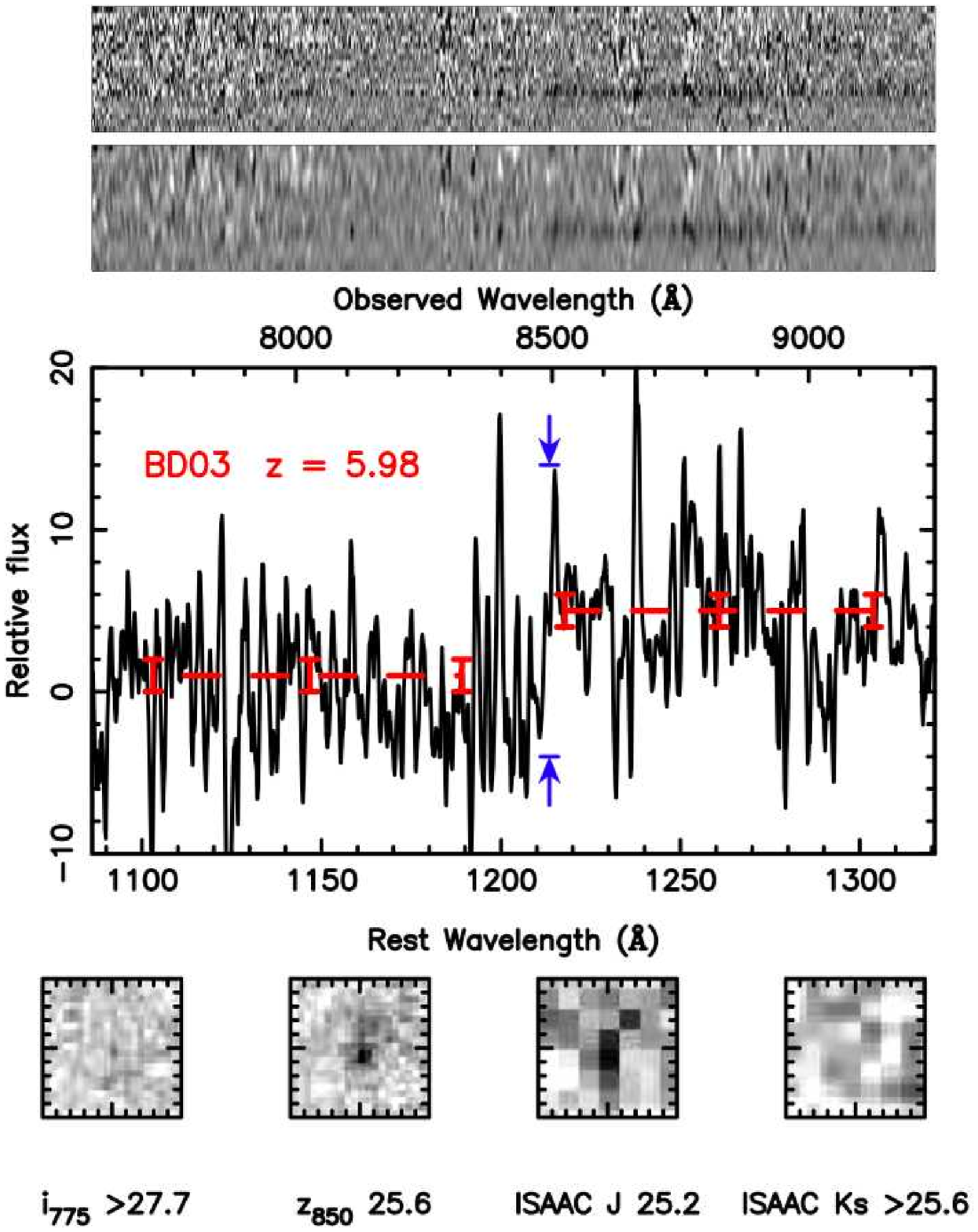}

\caption{\em FORS2 spectra and ACS/ISAAC imaging of BD03 (z = 5.98 \
$\pm$ \ 0.1). The upper two dimensional spectrum is unsmoothed, while
the lower one is smoothed with a $\sigma$=1 pixel Gaussian.  The
extracted one-dimensional spectrum is smoothed with a 5 pixel boxcar
filter.  From left to right, the lower four images are $i_{775}$,
$z_{850}$, \textit{J}, and \textit{K}.  The total exposure time is
22.3 hours.  The lower axis is the observed wavelength shifted to the
systematic redshift \textit{z} = 5.98.  Blue arrows indicate the
position of the continuum break.  The dotted lines represent the
wavelength region used to fit the continuum near the break, where
error bars represent the 1 $\sigma$ error.
\label{620}}

\end{center}
\end{figure}

\begin{figure}
\begin{center}
\figurenum{5}
\includegraphics[width=6in,  scale=0.8]{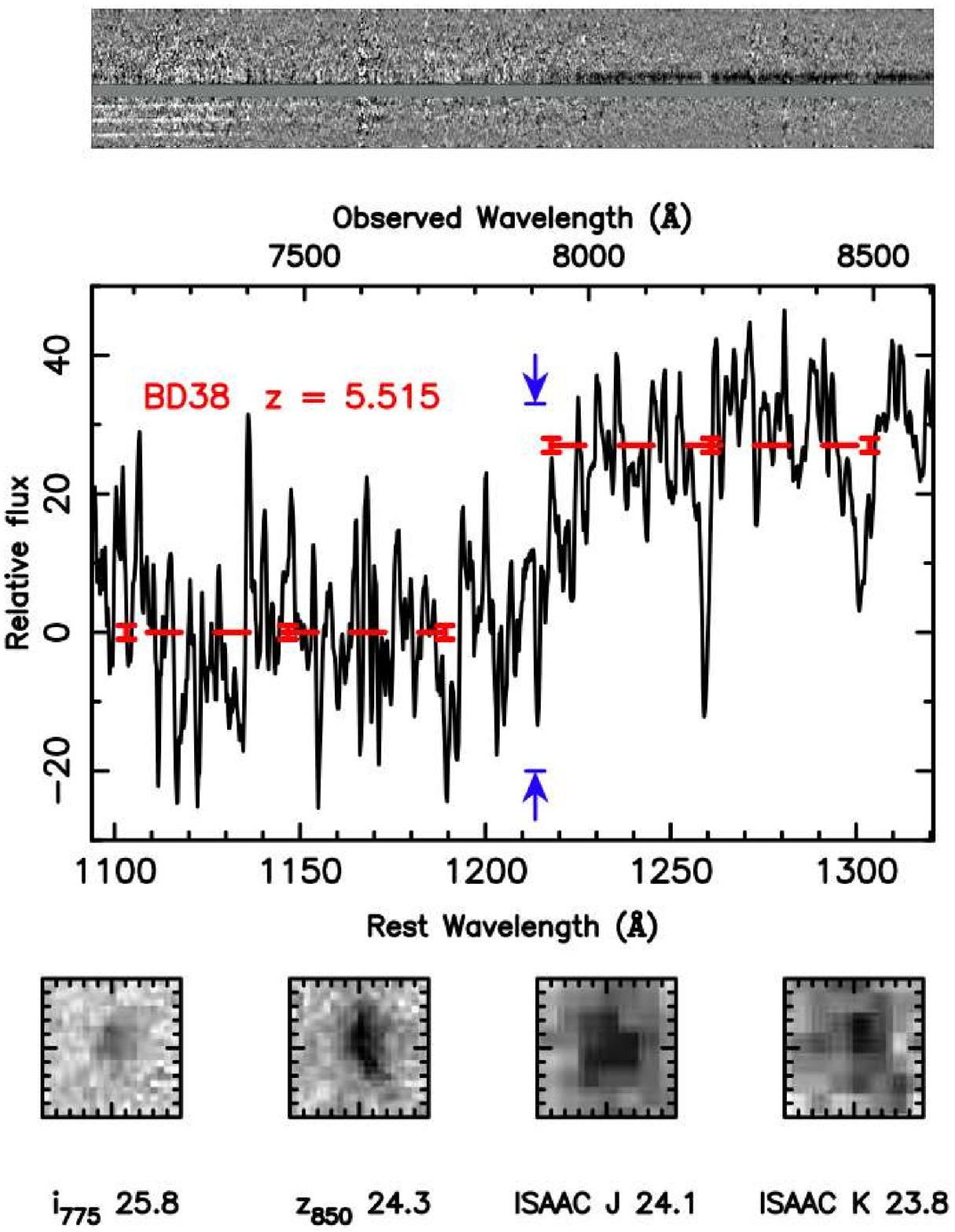}

\caption{\em FORS2 spectra, ACS and ISAAC imaging of BD38:
two-dimensional spectrum and extracted one-dimensional spectrum
smoothed with a 5 pixel boxcar filter.  From left to right, the lower
four images are $i_{775}$, $z_{850}$, \textit{J}, and \textit{K}.  The
total exposure time is 22.3 hours. The lower axis is the observed
wavelength shifted to the systematic redshift \textit{z} = 5.515.
Blue arrows indicate the position of the continuum break.  The dotted
lines represent the wavelength region used to fit the continuum near
the break, where error bars represent 1 $\sigma$ error.  A more
complete spectrum is shown in \citet{DowHygelund2005b}.
\label{4202}}

\end{center}
\end{figure}

\end{document}